\documentclass[reprint,twocolumn,superscriptaddress,showpacs,floatfix]{revtex4-1}
\usepackage{epsfig}
\usepackage{amsmath}
\usepackage{exscale}
\usepackage{array}
\usepackage{units}
\usepackage{amssymb}
\usepackage{setspace}
\usepackage{bbm}
\usepackage{color}
\usepackage{color}
\definecolor{lightgrey}{rgb}{0.89,0.89,0.89}
\definecolor{orange}{rgb}{0.89,0.39,0.09}

\begin{document}

\title{Ionization potentials and electron affinities  from the extended Koopmans' theorem in self-consistent Green's function theory}

\author{Alicia Rae Welden}
\affiliation{Department of Chemistry, University of Michigan, Ann Arbor, Michigan 48109, USA}
\author{Jordan J. Phillips}
\email{philljj@umich.edu}
\affiliation{Department of Chemistry, University of Michigan, Ann Arbor, Michigan 48109, USA}
\author{Dominika Zgid}
\affiliation{Department of Chemistry, University of Michigan, Ann Arbor, Michigan 48109, USA}

\begin{abstract}

One-body Green's function theories implemented on  the real  frequency axis  offer a natural formalism for the unbiased theoretical determination of quasiparticle spectra in molecules and solids. Self-consistent Green's function methods employing the imaginary axis formalism on the other hand can benefit from the iterative implicit resummation of higher order diagrams that are not included when only the first iteration is performed.  
Unfortunately, the imaginary axis Green's function  does not give direct access to the desired quasiparticle spectra, which undermines its utility. To this end we investigate how reliably one can calculate quasiparticle spectra  from the Extended Koopmans' Theorem (EKT) applied to the imaginary time Green's function in a second order approximation (GF2). We find that EKT in conjunction with GF2 yields IPs and EAs that systematically underestimate experimental and accurate coupled-cluster reference values for a variety of  molecules and atoms. This establishes that the EKT allows one to utilize the computational advantages of an imaginary axis implementation, while still being able to acquire real axis spectral properties. Because the EKT requires negligible computational effort, and  can be used with a Green's function from any level of theory, we conclude that it is a potentially very useful tool for the systematic study of quasiparticle spectra in realistic systems.


\end{abstract}
\maketitle


\section{Introduction}

Finding practical and principled methods to numerically solve the many-electron Schr\"{o}dinger equation for realistic chemical systems is a substantial problem that has been attacked by the scientific community for more than five decades now. Though diverse in approach, the bulk of these methods can be classified as either being based on the single-particle density $\rho(\mathbf{r})$ or the many-body wavefunction $\Psi$. Kohn-Sham density functional theory (DFT)\cite{PhysRev.136.B864,*PhysRev.140.A1133,DFTbookParr} and the coupled cluster theory (CC)\cite{CoupledClusterTheory_nucphys_1960,CC_Cizek,CCBartlett07} are two remarkably successful examples of these former and latter categories. 
Despite their merits and widespread use, by now it's become well established these classes of methods have their respective limitations that are unlikely to be overcome in the near future. For instance  CC theory, though being the gold standard for weakly correlated systems, cannot successfully account for multi-reference effects present in strongly correlated molecules and solids, and its computational scaling  seems to limit  use for large systems.
DFT in contrast is computationally affordable\cite{ScusScale1999}, but is plagued by a lack of systematic approaches to improve functionals\cite{Burke_Perspective_jcp_2012}, as well as the fact that many approximate functionals can give a frustratingly  non-uniform performance across different systems and properties. This problem is exemplified for metal oxides, where even within the same system different DFT approximations are  required for different properties\cite{EffectofXC_Gerson_prb_2000,Moreira:2002ad}. 
Furthermore,  as spectral properties are concerned, even with the exact functional the Kohn-Sham (KS) gap cannot provide a theoretically sound approximation to the fundamental gap\cite{Perdew_KS_Orbitals_1983,Perdew_BandGapProblem_ijqc_1985,OneDDMRGDFT_Burke_prl_2012,KSgap_Fundgap_Optgap_PCCP_2013}.

Approaches based on the self-consistent single-particle Green's function, $G(\omega)$, are an attractive alternative with strengths and weaknesses that are complementary to density and wavefunction-based methods.
 Similar to DFT, Green's function methods can be implemented in a blackbox manner in the atomic orbital (AO) basis\cite{GoWo_Koval_jcp_2011,scGWprb2013Scheffler,gf2_paper_2014,FullySCGW_Koval_prb_2014}, and typically will feature lower order polynomial  scaling    than wavefunction-based methods such as CCSD (coupled cluster singles doubles). Additionally, Green's function methods offer a natural language for embedding approaches since the Green's function can be partitioned among subsystems (similar to the density). 
 
Because of the need to evaluate $G(\omega_n)$ on a numerical grid, $\omega_n$, Green's function based approaches can be broadly grouped into two classes:

{\em Greens functions on the imaginary (Matsubara grid) axis} describe a grand canonical ensemble and can be used to 
calculate temperature dependent quantities such as the free energy, specific heat, etc. A Green's function of the imaginary axis is a smooth function, with only a single pole near zero that is not accessed for any finite temperature larger than zero. Consequently, the imaginary axis formulation is a natural choice for any self-consistent approach where the Green's function is expressed as a functional of the self-energy, $G[\Sigma(\omega)]$, and where at self-consistency infinite classes of diagrams can be included due to the implicit resummation. 
A result of this diagrammatic resummation is that these approaches\cite{gf2_paper_2014,Dahlenjcp2005,levelsofscGWjcp2009Dahlen,scGWprb2013Scheffler} give finite results in strongly correlated cases when other methods such as truncated CC or MP2 (second order M\o ller-Plesset\cite{MollerPlesset1934}) would diverge pathologically.


{\em Greens functions on the real frequency axis} are more commonly employed in quantum chemistry for zero-temperature calculations. The single-particle Green's function on the real axis has multiple poles which correspond to ionization potential (IP) and electron affinity (EA) peaks~\cite{HedinGWpra1965,IPsEAs_2ndOrderEPT_Jorgensen_jcp_1979,Ortiz_EPTofEAs_cpl_1981,ADC_pra_1983,OnGfAndTheirApplicationsCederbaum1990,Massidda_MnO_GW_prl_1995,*Massidda_MetalOxide_GW_prb_1997,BandGapProblem_scGW_KuWei_prl_2002,Fully_scGW_Rostgaard_prb_2010,Ortiz_Review_2013} in the photoelectron spectrum. The two-particle real-axis Green's function is capable of describing optical/neutral excitations.
Performing the calculation self-consistently  (where $G[\Sigma(\omega)]$ is evaluated as a functional of $\Sigma(\omega)$) is notoriously difficult on the real axis, because the series of poles in the real-axis Green's function requires a non-uniform grid that can change between  iterations.

The real and imaginary axis formalisms should be treated as complementary but requiring the development of different tools in order to evaluate accurate Green's functions. The interesting question that arises is if one can use an approach that has ``the best of both worlds"; that is, to calculate the Green's function on the imaginary axis in order to take advantage of the efficient self-consistency, and to subsequently employ the imaginary axis solution to calculate spectra on the real axis.

One of the routes to obtaining a real axis Green's function from the imaginary axis data is through the process of analytic continuation~\cite{Analytic_cont1, Analytic_cont2, Analytic_cont3}. This procedure is known to suffer from several problems, namely it cannot recover sharp spectral features, frequently is problematic in recovering fundamental gap edges, and is very sensitive to the initial imaginary axis data.

Here, we attempt to examine an alternative answer to this problem.  We will investigate how accurately and reliably one can calculate quasiparticle spectra from the extended Koopman's theorem (EKT)\cite{EKT_Original_jcp_1975,*EKT_Original_II,EKT_Pickup,KatrielDavidson_EKT_pnas_1980,EKT_IJQC,EKT_Morrison,Olsen_EKT_jcp_1993,AyersEKT_jcp_1995,IPandEAfromEKT_Cioslowski_jcp_1997,Olsen_PertExpansion_EKT_cpl_1998,EKT_Cioslowski2001,Ernzerhof_Validity_of_EKT_jctc_2009,Bozkaya_EKT_jcp_2013,Bozkaya_EKT_jctc_2014} starting from  self-consistent Green's function many-body theory\cite{gf2_paper_2014,Dahlenjcp2005,levelsofscGWjcp2009Dahlen} in an imaginary axis implementation. The EKT is valuable because it allows one to obtain, in principle, both IPs and EAs 
 from a single electronic structure calculation on the neutral system. Similar to population analysis\cite{Reed_NPA_jcp_1985}, the actual EKT procedure itself takes place in a simple post-calculation analysis, utilizing the density matrix and other quantities obtained by the preceding correlated method. As a result the EKT can be implemented in a blackbox manner, requires only a negligible fraction of time, and is not tied to any particular level of theory. Despite its simplicity, most of the efforts so far seem to have focused on its application for calculating IPs only\cite{EKT_Original_jcp_1975,*EKT_Original_II,EKT_Pickup,EKT_Morrison,Morrison_EKT_MCSCF_jcc_1992,IPandEAfromEKT_Cioslowski_jcp_1997,Pernal_IP_EKT_cpl_2005,Dahlenjcp2005,levelsofscGWjcp2009Dahlen,Bozkaya_EKT_jcp_2013}. While an interesting benchmark of the EKT for EAs has appeared recently\cite{Bozkaya_EKT_jctc_2014}, in that work Bozkaya actually obtained the EA \emph{indirectly} by calculating the IP of the anion. In contrast, in this work we will calculate   IPs and EAs  from the neutral system via the EKT. As far as we are aware, EAs  typically have not been  calculated in this manner.  
 
 While the EKT has usually been formulated in terms of a generalized Fockian\cite{IPandEAfromEKT_Cioslowski_jcp_1997} that is evaluated using methods such as configuration interaction (CI) or MP2, in this work we approach EKT with the machinery of second order Green's function theory (GF2)~\cite{gf2_paper_2014,Dahlenjcp2005}.
 As shown by Dahlen, Stan, and van Leeuwen\cite{Dahlenjcp2005,levelsofscGWjcp2009Dahlen}, via EKT it is possible to calculate both  IPs and EAs from the imaginary time Green's function  of the neutral system alone. This is potentially very useful for studying the spectral properties of extended systems, because it would circumvent the numerically ill defined step of analytic continuation to the real axis. While their initial results for IPs obtained with GF2 and GW were promising, here we intend to examine GF2's performance for IPs and EAs for a  wider group of atoms and small molecules. 

\section{Theoretical calculations of ionization potentials and electron affinities}

To make this work self-contained, here we briefly review some of the different strategies that have been utilized to calculate IPs, EAs,  and the fundamental gap $E_{\scriptstyle g}=\textnormal{IP}-\textnormal{EA}$, using wavefunction theory, DFT, many-body theory, and combinations thereof. Naively, the simplest strategy for obtaining these quantities would be by energy differences of the neutral and charged cation/anion systems. However this is fundamentally problematic for periodic boundary-conditions (PBC) calculations of materials in the solid-state, where it would imply the presence of an infinite amount of unbalanced charge in the crystal.  Even for finite systems, careful early  CI studies\cite{Sasaki_EA_PRA_1974_i,*Sasaki_EA_PRA_1974_ii,Feller_Davison_EAofOxy_jcp_1989,Dunning_ElecAffinity_jcp_1992} found that obtaining accurate EAs from energy-differences was particularly challenging because the correlation energy of the anion could converge appreciably slower than that of the neutral system. 
Additionally, methods with significant amounts of self-interaction error (such as approximate DFT) may not be able to bind some anions at all\cite{CommentAnion1996,ConcerningAnion1996,JensenAnion2010}. 

For these reasons, significant interest has been placed in  obtaining IPs and EAs \emph{directly} from a single calculation on the neutral system. An exemplar for this is offered by the equation of motion coupled cluster theory (EOM-CC)\cite{EOMCC_Emrich_nucphysa_1981,EOMCC_Sekino_ijqc_1984,EOMCC_Geertsen_cpl_1989,EOMCC_Stanton_jcp_1993,EOM_Krylov_arpc_2008} for electron attachment and removal\cite{Marcel_EA_EOM_CC_jcp_1995,Marcel_SimTrans_EOM_CC_IP_EA_EE_jcp_1997,Musial_IP,Musial_EA,Musial_organic,Kamiya_Hirata_EA_EOM_CC_jcp_2007,Musial}. EOM-CC is a generalization of the original coupled cluster theory to charged and neutral excitations, and thus largely inherits the advantages and disadvantages of the CC method: if a single determinant description is valid, and the calculation not prohibitively expensive, one can expect very accurate results. 
 Closely related to  IP/EA-EOM-CC is the coupled cluster Green's function method of Nooijen and Snijders\cite{Marcel_CC_GreensFunction_ijqc_1992,*Marcel_CC_GreensFunction_ijqc_1993,*Marcel_gf2_CC_IP_jcp_1995}, that again will bear some of the  advantages and disadvantages of the underlying CC theory.

 In the same spirit,  Green's function (also called electron propagator) methods such as the n$^{th}$-order Algebraic Diagrammatic Approximation (ADC{\footnotesize(n)})\cite{ADCn,ADC_pra_1983,ADC_pra_1989} and various self-energy approximations\cite{OrtizThreeApproximations_ijqc_2010} have found regular use  for the accurate, direct determination of IPs and EAs in finite systems\cite{DirectIPs_Cederbaum_1973,Ortiz_EPTofEAs_cpl_1981,Ortiz_Partial4thOrder_jcp_1988,Ortiz_NonDiagPartial4th_ijqc_1989,OnGfAndTheirApplicationsCederbaum1990,Ortiz_Partial3rdOrder_jcp_1996,Ortiz_NonDiagRenorm_3rdOrder_jcp_1998,CarbonCluster_ADC_jcp_1999,Ortiz_EPT_jcp_2004,NonDysonADC_jcp_2005,HowMuchDoublePolyene_ADC_2006,Cederbaum_Water_DoubleIP_jcp_2006}. Typically the underlying scheme in these methods is the iterative diagonalization of the self-energy on the real axis in the molecular orbital (MO) basis, $\mathbf{\Sigma}(\lambda_{i})\mathbf{c}_{i}=\lambda_{i}\mathbf{c}_{i}$, until one converges to a given IP, $\lambda_{i}$. As such, this is quite distinct from the fully self-consistent Green's function implementations 
 in the local AO basis 
 that have  appeared recently\cite{scGWprb2013Scheffler,FullySCGW_Koval_prb_2014,gf2_paper_2014}  and that we are considering presently in this work.

 A  theoretically not fully justified however computationally cheap strategy for directly obtaining quasiparticle spectra would be to simply perform a DFT calculation on the neutral system, and then interpret the resulting KS eigenvalues as IPs  and EAs.
Unfortunately, as previously mentioned there is no rigorous theoretical grounding for such a procedure\cite{Perdew_KS_Orbitals_1983,Perdew_BandGapProblem_ijqc_1985,AccXCPotSi_Discontinuity_Godby_prl_1986,DFTMBPT_BandGap_Gruning_jcp_2006,KSgap_Fundgap_Optgap_PCCP_2013}: even if the exact functional were used, while the   highest occupied KS orbital eigenvalue would be the negative of the ionization potential, the KS gap $E^{\scriptscriptstyle KS}_{\scriptstyle g}$ would still differ from the exact fundamental gap $E_{\scriptstyle g}$  by an amount equalling the derivative discontinuity of the exchange-correlation (XC) potential, $E_{\scriptstyle g}=E^{\scriptscriptstyle KS}_{\scriptstyle g}+\Delta_{\scriptscriptstyle XC}$. As shown recently with stretched hydrogen chains, when strong correlation is present the derivative-discontinuity $\Delta_{\scriptscriptstyle XC}$ can actually become the dominant contribution to $E_{\scriptstyle g}$\cite{OneDDMRGDFT_Burke_prl_2012}.
If one disregards this principled objection,  in practice the KS gap still  underpredicts the fundamental gap significantly, which is partly a consequence of the fractional-charge error in approximate density functionals\cite{SIEcPerdewZunger1981,Perdew1982,SpuriousFrac2006,ManyeSIEYang2006,FractionalBandGap_Cohen_prb_2008}.

Because Hartree-Fock and KS DFT tend to have opposing fractional-charge errors, 
 these methods will typically  overestimate and underestimate the fundamental gap, respectively. For this reason there will almost always be a system-dependent empirical hybrid functional that will describe the fundamental gap well\cite{EffectofXC_Gerson_prb_2000,Moreira:2002ad,HSE_ParameterSpace_Moussa_jcp_2012} when calculated in the Generalized Kohn-Sham (GKS)\cite{GeneralizedKohnSham_BandGap_Levy_prb_1996} formalism as band energy-differences. From this standpoint it's understandable why the HSE functional\cite{Heyd:2003db,*Heyd:2004ud} has enjoyed so much success for  many insulators and semiconductors\cite{EnergyBandGaps_HSE_Heyd_Peralta_jcp_2005,BandGapsHSE_vero_barone_jcp_2008,AccurateSolidsWithHSE_Henderson_pss_2011}, though it's been remarked the HSE gap will tend to match the optical gap better than the fundamental gap\cite{BandGapsHSE_vero_barone_jcp_2008,AccurateSolidsWithHSE_Henderson_pss_2011}. 

Because of these fundamental issues with the KS eigenvalues, a very commonly used strategy has been to  perform ``one-shot'' corrections to the DFT spectra with the $G_{0}W_{0}$ approximation\cite{HedinGWpra1965,Hybertsen_GW_prb_1986,GW_Rubio_RevModPhys_2002}. From a purely pragmatic viewpoint the resulting quasiparticle spectra can be much improved with respect to experiment. However from a principled standpoint this is not entirely satisfying because it is highly dependent on the starting DFT solution which is functional dependent. Because of this, the $G_{0}W_{0}$ results can vary significantly depending on the combination of system and density functional used\cite{GWStartingPoint_Kresse_prv_2007,GWStartingPoint_prb_2011,GWStartingPoint_TiO2_Gap_Moussa_prb_2011,GWStartingPoint_Kronik_prb_2011,oneshotGWReference_Carter_pccp_2011,DFT_MetalOxide_BlasphemyPaper_pccp_2011,oneshotGWReference_Carter_prb_2012,GWBenchmark_prb_2012,GWStartingPoint_prb_2012,GWStartingPoint_jctc_2013}. To highlight one such example, for hematite ($\alpha$-Fe$_{2}$O$_{3}$) the $G_{0}W_{0}$ correction can yield  quasiparticle gaps ranging from 1.3, 4.0, to 4.5~eV depending on whether PBE\cite{PBE,*PBE1}, HSE\cite{Heyd:2003db,*Heyd:2004ud}, or PBE0\cite{PBE0} is used, respectively, which is contrasted with the experimentally determined gap of 2.6$\pm$0.4~eV\cite{oneshotGWReference_Carter_pccp_2011}.
Though controversy exists in whether self-consistency will overall improve or worsen results\cite{scGW_Holm_vonBarth_PRB_1998,scGW_Semiconductors_prl_1998,BandGapProblem_scGW_KuWei_prl_2002,levelsofscGWjcp2009Dahlen,Fully_scGW_Rostgaard_prb_2010,GW_DNA_selfconsistent_prb_2011,scGW_transport_prb_2011,scGW_Unified_Scheffler_prb_2012}, clearly fully self-consistent Green's function calculations are valuable because they are reference independent. We review one such self-consistent implementation next.

\section{GF2 Theory and Implementation}

The  real axis single-particle Green's function, $G(\omega)$, determines the expectation value of all single-particle observables, in addition to the spectral density of states, IPs, and EAs. Unfortunately, calculating $G(\omega)$ exactly for a large system is not any more feasible than  calculating the exact wavefunction $\Psi$.  Nonetheless \emph{we can} calculate the Green's function of a non-interacting system, $G_{0}(\omega)$, very easily, and then correct it for the missing many-body correlation effects in a systematic way via the Dyson equation.
 Given some  $G_0(\omega)$, the exact  $G(\omega)$ can be obtained by expanding in terms of the proper self-energy, $\Sigma(\omega)$, and analytically summing to yield the Dyson equation

\begin{equation}
\begin{split}
G(\omega)=G_{0}(\omega)+G_{0}(\omega)\Sigma(\omega)G_{0}(\omega)\\+G_{0}(\omega)\Sigma(\omega)G_{0}(\omega)\Sigma(\omega)G_{0}(\omega)+\cdots \\
=G_{0}(\omega)\biggr(\sum_{n}\bigr(\Sigma(\omega)G_{0}(\omega)\bigr)^{n}\biggr)\\ 
=\bigr[G_{0}(\omega)^{-1}-\Sigma(\omega)\bigr]^{-1}\\
\end{split}
\label{eq:DysonSum}
\end{equation}

\begin{figure}
\includegraphics[width=8.5cm]{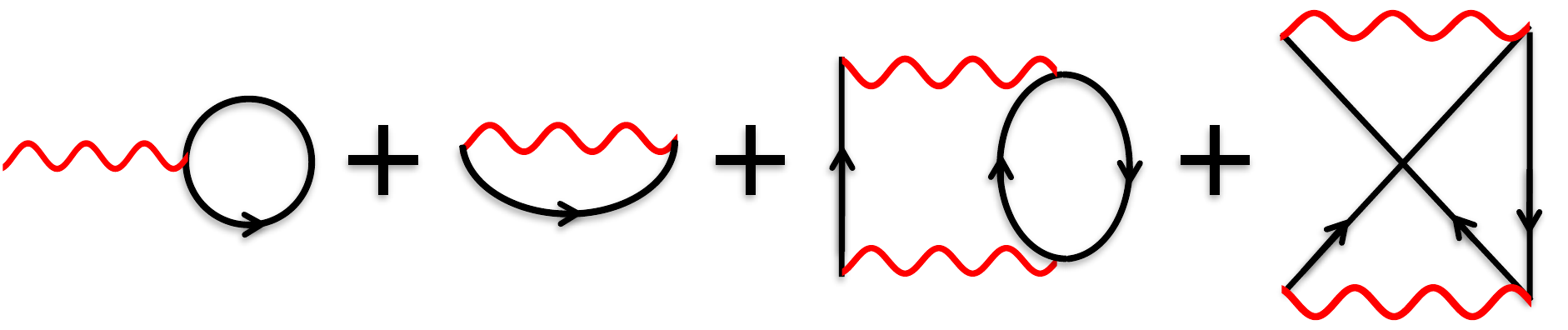}
\caption{Feynman diagrams included in the GF2 self-energy. A black arrow line represents a Green's function, while a red squiggle represents a two-electron integral. The first two diagrams are the frequency-independent Hartree and exchange terms, and are included in the Fock matrix. The next two diagrams are the frequency-dependent pair-bubble and second order exchange terms, and represent the second order correlation effects covered by $\mathbf{\Sigma}(\omega)$.}
\label{fig:diagrams}
\end{figure}

\noindent The self-energy, $\Sigma(\omega)$, is a frequency-dependent single-particle potential that encompasses all of the exchange-correlation (XC) effects of the many-body system. Analogous to the KS potential $V_{\scriptscriptstyle XC}$, one could think of $\Sigma(\omega)$ as being the XC potential that connects the Green's function of the non-interacting system to the Green's function of the fully interacting system. However, it is important to remember that  $\Sigma(\omega)$ is dynamic, nonlocal, and orbital-dependent, while $V_{\scriptscriptstyle XC}$ in approximate DFT is typically static, ``semilocal'', and density-dependent. In principle the exact $\Sigma(\omega)$ can be expanded diagrammatically in  $G(\omega)$\cite{Mattuck_Feynman,fetter2003quantum}. In a practical implementation, one chooses a subset of diagrams that can be evaluated in a computationally tractable manner. The resulting self-energy can then be written as an approximate functional of the Green's function, $\Sigma[G(\omega)]$, yielding a self-consistent set of equations. 
In this work we investigate the second order approximation (GF2), which includes all diagrams to second order and is shown in Fig.~\ref{fig:diagrams}. 

To take advantage  of the easy to converge self-consistency procedure on the imaginary axis, the Green's function can be written in a non-orthogonal atomic-orbital (AO) basis  as

\begin{equation}
\mathbf{G}(\omega)=\bigr[(\mu+\omega)\mathbf{S}-\mathbf{F}-\mathbf{\Sigma}(\omega)\bigr]^{-1} ~~,
\label{eq:GofOmega}
\end{equation}

\noindent where $\mathbf{S}$ and $\mathbf{F}$ are the overlap and Fock matrices, $\mu$ is the chemical potential, $\omega$ is an imaginary frequency, and $\mathbf{\Sigma}(\omega)$ is the aforementioned frequency-dependent self-energy within the GF2 approximation containing second order diagrams from Fig.~\ref{fig:diagrams}. We use a uniform grid of imaginary Matsubara frequencies $\omega_{n}=(2n+1)i\pi/\beta$,
 with a power mesh imaginary time grid\cite{ALPS} running on the interval $0<\tau<\beta$, where $\beta$ is the inverse temperature. We choose to build the Green's function on the frequency axis because of the simplicity of Eq.~\ref{eq:GofOmega}, in contrast to the expression for $\mathbf{G}(\tau)$ which is more cumbersome and requires integrations over $\tau$ points\cite{Dahlenjcp2005}. Once built, the imaginary frequency Green's function can be fast Fourier transformed (FFT) to the imaginary time domain, $\mathbf{G}(\tau)$.  The correlated density matrix $\mathbf{P}$ then is  evaluated as

\begin{equation}
\mathbf{P}=-2\mathbf{G}({\scriptstyle \tau=\beta}) ~.
\label{eq:PofG}
\end{equation}

\noindent Provided with $\mathbf{P}$, the correlated Fock matrix is built by

\begin{equation}
F_{ij}=h_{ij}+\underset{kl}{\sum}P_{kl}(\textnormal{v}_{ijlk}-\frac{1}{2}\textnormal{v}_{iklj})~,
\label{eq:FockMatrix}
\end{equation}

\noindent where $h_{ij}$ and $\textnormal{v}_{ijlk}$ are one and two-electron integrals in the AO basis. Note that the (frequency-independent) first-order self-energy is already covered by the Hartree-Fock mean-field, $\Sigma_{\infty}={\sum_{kl}}P_{kl}(\textnormal{v}_{ijlk}-\frac{1}{2}\textnormal{v}_{iklj})$. Finally, the (frequency dependent) second order self energy can be built in the time-domain as

\begin{equation}
\begin{split}
\Sigma_{ij}(\tau)=-\underset{klmnpq}{\sum}G_{kl}(\tau)G_{mn}(\tau)G_{pq}(-\tau)\\
\times\textnormal{v}_{imqk}\bigr(2\textnormal{v}_{lpnj}-\textnormal{v}_{nplj}\bigr)~,
\end{split}
\label{eq:Sigma}
\end{equation}

\noindent and then FFT to the frequency domain. It is simpler to build $\mathbf{\Sigma}$ on the time axis, because in the $\tau$ domain the self-energy factorizes into simple products of Green's functions, whereas in the $\omega$ domain it requires integrations of Green's functions over frequencies. Furnished with an updated $\mathbf{F}$ and $\mathbf{\Sigma}(\omega)$, we can return to Eq.~\ref{eq:GofOmega} and rebuild $\mathbf{G}(\omega)$. Taken altogether Eq.~\ref{eq:GofOmega}, \ref{eq:PofG}, \ref{eq:FockMatrix}, and \ref{eq:Sigma} present a self-consistent procedure for solving the Dyson equation in a second order approximation to the self energy. To initiate the self-consistency an approximate zeroth order Green's function, $\mathbf{G}_{0}(\omega)$, is necessary, which practically can be supplied by DFT or Hartree-Fock (HF) calculations. In this work we use an initial HF Green's function (i.e. $\mu\approx\frac{1}{2}(\epsilon_{\scriptscriptstyle HOMO}+\epsilon_{\scriptscriptstyle LUMO})$, $\mathbf{F}=\mathbf{F}_{\scriptstyle HF}$, and $\mathbf{\Sigma}(\omega)=\mathbf{0}$) generated via output from the Dalton electronic structure program\cite{Dalton}. At self-consistency $\mathbf{G}(\omega)$ will not depend on the starting-reference\cite{FullySCGW_Koval_prb_2014}, though practically some initial guesses might be better than others for converging rapidly.
As a final note, it should be understood that the Green's function depends on $\mu$, and therefore $\mathbf{P}$ implicitly depends on $\mu$ as well. This means the chemical potential will need to be adjusted from iteration-to-iteration to maintain the correct electron number.

For purposes of comparison,  we will also consider a non self-consistent Green's function obtained from the first iteration of the Dyson equation, given by

\begin{equation}
\mathbf{G}_{1}(\omega)=\bigr[(\mu+\omega)\mathbf{S}-\mathbf{F}_{HF}-\mathbf{\Sigma}[G_0(\omega)]\bigr]^{-1} ~~.
\label{eq:GofOmegaOneIter}
\end{equation}

\noindent Here $\mathbf{\Sigma[G_0(\omega)]}$ is simply the self-energy obtained when $\mathbf{G}_{0}$, which in our case is $\mathbf{G}_{HF}$, is inserted into Eq.~\ref{eq:Sigma}, and $\mu$ is set so that $\mathbf{G}_{0}(\omega)$ has good particle number. For conciseness, we will refer to this simply as G$_{0}$F2, in analogy to  $G_{0}W_{0}$. 
Since this Green's function is not self-consistent, it will carry a starting reference dependence. 

Since the Green's function obtained by self-consistent or non-self-consistent GF2 is expressed on an imaginary grid, we aim to employ the Extended Koopman's Theorem (EKT)\cite{EKT_Original_jcp_1975,*EKT_Original_II,EKT_Pickup,EKT_IJQC,EKT_Morrison,AyersEKT_jcp_1995,Bozkaya_EKT_jcp_2013,Bozkaya_EKT_jctc_2014} to obtain 
 ionization potentials and electron affinities, which are real axis quantities and can be used to produce the spectral density of states $A(\omega)$ expressed as $A(\omega)=-\frac{1}{\pi}\textnormal{Tr}[\textnormal{Im\,}\mathbf{G}(\omega)\mathbf{S}]$.  We give a brief discussion of the EKT theory and implementation next.


\subsection{Extended Koopmans' Theorem}

Given  a system with  Hamiltonian $\hat{H}$ and $N$ electron state $|{\scriptstyle N}\rangle$ satisfying $\hat{H}|{\scriptstyle N}\rangle=E|{\scriptstyle N}\rangle$, by using  second-quantized operators, $\hat{a}|{\scriptstyle N}\rangle=|{\scriptstyle N-1}\rangle$, $\hat{a}^{\dagger}|{\scriptstyle N}\rangle=|{\scriptstyle N+1}\rangle$, the energies of the anion, neutral, and cation states can be expressed as

\begin{equation}\begin{split}
E_{{\scriptstyle N+1}}=\langle{\scriptstyle N}|\hat{a}\,\hat{H}\,\hat{a}^{\dagger}|{\scriptstyle N}\rangle ~, \\
E_{{\scriptstyle N}}=\langle{\scriptstyle N}|\hat{H}|{\scriptstyle N}\rangle ~,\\
E_{{\scriptstyle N-1}}=\langle{\scriptstyle N}|\hat{a}^{\dagger}\,\hat{H}\,\hat{a}|{\scriptstyle N}\rangle ~,
\end{split}\label{eq:AnionNeutralCation}\end{equation}

\noindent respectively. It should be understoood these operators are expanded in a basis, $\hat{a}=\sum_{i}c_{i}\hat{\phi}_{i}$, with the expansion coefficients $c_{i}$ chosen so that the anion (cation) state remains normalized\cite{AyersEKT_jcp_1995}. Provided with Eq.~\ref{eq:AnionNeutralCation}  the ionization potential ($I$) and electron affinity ($A$) can be defined as

\begin{equation}\begin{split}
I=E_{\scriptstyle N-1}-E_{\scriptstyle N} = -\langle{\scriptstyle N}|\hat{a}^{\dagger}[\hat{a},\hat{H}]|{\scriptstyle N}\rangle ~, \\
A=E_{\scriptstyle N}-E_{\scriptstyle N+1} =\langle{\scriptstyle N}|\hat{a}[\hat{a}^{\dagger},\hat{H}]|{\scriptstyle N}\rangle ~.
\label{eq:IPEA}
\end{split}\end{equation}

\noindent A Lagrangian for $I$ and $A$ can now be constructed, given by

\begin{equation}\begin{split}
\mathcal{L}_{I}=-\langle{\scriptstyle N}|\hat{a}^{\dagger}[\hat{a},\hat{H}]|{\scriptstyle N}\rangle +\epsilon(\langle{\scriptstyle N}|\hat{a}^{\dagger}\hat{a}|{\scriptstyle N}\rangle-1) ~,  \\
\mathcal{L}_{A}=\langle{\scriptstyle N}|\hat{a}[\hat{a}^{\dagger},\hat{H}]|{\scriptstyle N}\rangle +\epsilon(\langle{\scriptstyle N}|\hat{a}\,\hat{a}^{\dagger}|{\scriptstyle N}\rangle-1) ~,
\end{split}\label{eq:HouseonLaGrange}\end{equation}

\noindent where the right-hand term constrains the cation/anion state to be normalized.  Expanding the operators in their basis,
 and exploiting that $\hat{a}\,\hat{a}^{\dagger}=\hat{1}-\hat{a}^{\dagger}\hat{a}$, the stationary solution $\delta\mathcal{L}/\delta{c_{i}^{\dagger}}=0$ of Eq.~\ref{eq:HouseonLaGrange} yields the  generalized eigenvalue problem

\begin{equation}\begin{split}
\mathbf{H}^{I}\mathbf{c}=\epsilon\mathbf{P}\mathbf{c} ~,\\
\mathbf{H}^{A}\mathbf{c}=\epsilon\mathbf{P}_{v}\mathbf{c} ~, 
\end{split}\label{eq:EigProb}\end{equation}

\noindent where 
%
$[\mathbf{H}^{I}]_{ij}\equiv \langle{\scriptstyle N}|\hat{\phi}_{i}^{\dagger}[\hat{\phi}_{j},\hat{H}]|{\scriptstyle N}\rangle$ and  
$[\mathbf{H}^{A}]_{ij}\equiv-\langle{\scriptstyle N}|\hat{\phi}_{i}[\hat{\phi}^{\dagger}_{j},\hat{H}]|{\scriptstyle N}\rangle$ are generalized Fock matrices, and $[\mathbf{P}]_{ij}\equiv\langle{\scriptstyle N}|\hat{\phi}_{i}^{\dagger}\hat{\phi}_{j}|{\scriptstyle N}\rangle$ is the density matrix.
$\mathbf{P}_{v}$ is the virtual (or ``hole'') density matrix, which is defined within the orthogonal L\"{o}wdin basis as $\mathbf{P}_{v}\equiv2\mathbf{I}-\mathbf{P}$, where $\mathbf{I}$ is the identity matrix and the factor of two accounts for double occupation in our spin-restricted formalism, or equivalently in terms of Green's functions in the L\"{o}wdin basis as $\mathbf{P}_{v}=-2\mathbf{G}({\scriptstyle \tau=0_+})$.

For practical calculations we want to connect Eq.~\ref{eq:EigProb} to Green's function many-body theory in the following way: From the definition of the time-dependent single-particle Green's function\cite{fetter2003quantum}, $G(\tau)$, one can show

\begin{equation}
\begin{split}
\lim_{\tau\rightarrow 0}\dfrac{\partial G(\tau)}{\partial\tau}=
\Biggr\{ \begin{matrix}
\langle N| \hat{a}[\hat{a}^{\dagger},\hat{H}]|N\rangle ~,& \tau>0\\
 \langle N|\hat{a}^{\dagger}[\hat{a},\hat{H}]|N\rangle ~,& \tau<0
\end{matrix}
\end{split}
\label{eq:TauG}
\end{equation}

\noindent For compactness we  simply write these two possibilities as $\partial_{\tau}G(\tau)|_{0^{+}}$ and $\partial_{\tau}G(\tau)|_{0^{-}}$.  Eq.~\ref{eq:TauG} has this form because of the discontinuity in the Green's function at $\tau=0$, and furthermore $G(\tau)=-G(\tau+\beta)$, which means that $\partial_{\tau}G(\tau)|_{0^{-}}=-\partial_{\tau}G(\tau)|_{\beta}$. The important point of Eq.~\ref{eq:TauG} is  the generalized Fockians appearing in the standard EKT Eq.~\ref{eq:EigProb} can be replaced with time-derivatives of the Green's function on the imaginary-domain in the AO basis. Introducing a matrix representation for the Green's function in this basis, $\mathbf{G}(\tau)$,  performing the transformation $\mathbf{c}=\mathbf{P}^{-1/2}\mathbf{c}'$ (or  $\mathbf{c}=\mathbf{P}_{v}^{-1/2}\mathbf{c}'$) and multiplying on the left with $\mathbf{P}^{-1/2}$ (or $\mathbf{P}^{-1/2}_{v}$) results in

\begin{equation}
\begin{split}
\Delta_{-}\mathbf{c}'=\epsilon_{I}\mathbf{c}' ~,~ \epsilon_{I}=I+\mu\\
\Delta_{+}\mathbf{c}'=\epsilon_{A}\mathbf{c}' ~,~ \epsilon_{A}=A+\mu\\
\Delta_{-}=2\mathbf{P}^{-1/2}\partial_{\tau}\mathbf{G}(\tau)|_{0^{-}}\mathbf{P}^{-1/2} \\
\Delta_{+}=-2\mathbf{P}^{-1/2}_{v}\partial_{\tau}\mathbf{G}(\tau)|_{0^{+}}\mathbf{P}^{-1/2}_{v}
\end{split}
\label{eq:GreenIPsEAs}
\end{equation}

\noindent where $\mathbf{P}=-2\mathbf{G}({\scriptstyle \tau=\beta})$, and $\mathbf{P}_{v}=-2\mathbf{G}({\scriptstyle \tau=0_+})$. The factor of two in Eq.~\ref{eq:GreenIPsEAs}  accounts for double occupation in our spin-restricted formalism. We emphasize  Eq.~\ref{eq:GreenIPsEAs} assumes that $\mathbf{G}(\tau)$ and $\mathbf{P}$ have been pre-transformed to the L\"{o}wdin basis.  Eq.~\ref{eq:GreenIPsEAs} shows that diagonalization of the $\Delta_{-}$ and $\Delta_{+}$ matrices gives eigenvalues which, after subtracting out the chemical potential $\mu$, yield the ionization potentials and electron affinities respectively. For example,  if the Hartree-Fock Green's functions were inserted in Eq.~\ref{eq:GreenIPsEAs} then this would yield simply the Koopman's theorem IPs and EAs (in fact this is a good way to check the accuracy of one's  grid). Conceptually, the form of Eq.~\ref{eq:GreenIPsEAs} can be understood by realizing that $\mathbf{G}(\tau)$ for $\tau$ near $\beta$ describes the particle distribution of the system and consequently electron removal, whereas for $\tau$ near $0_+$ it contains information on the hole distribution and therefore electron attachment.

One subtlety is that diagonalizing $\Delta_{\pm}$ will of course yield as many eigenvalues as there are AO basis functions, $N_b$, yet only some of these eigenvalues may be physically meaningful as IPs or EAs. We find the simplest way to identify the correct eigenvalues is by the corresponding Dyson occupations

\begin{equation}\begin{split}
\mathbf{C}_{-}^{\dagger}\mathbf{P}\,\mathbf{C}_{-}=\mathbf{D}~~, \\
\mathbf{C}_{+}^{\dagger}\mathbf{P}_{v}\mathbf{C}_{+}=\mathbf{D}_{v} ~~.
\end{split}\end{equation}

\noindent Here $\mathbf{C}_{\pm}$ is the matrix of eigenvectors, $\mathbf{C}=\{\mathbf{c}^{'}_{1},\mathbf{c}^{'}_{2},\ldots\}$, obtained from diagonalizing $\Delta_{\pm}$. The diagonal elements of $\mathbf{D}$ ($\mathbf{D}_{v}$) correspond to occupations of Dyson orbitals for electron removal (attachment). As a consistency check one should find that $\textnormal{Tr}[\mathbf{D}]=N$, and  $\textnormal{Tr}[\mathbf{D}_{v}]=2N_{b}-N$. 
 Essentially the orbitals with large occupations (roughly speaking $1<[\mathbf{D}]_{ii}<2$) will indicate the IPs/EAs one is interested in.

As a final note, we stress again that the use of the Extended Koopmans' Theorem is not limited to GF2 or even Green's function methods, and has been employed with a variety of methods at different levels of theory in the past\cite{EKT_Original_jcp_1975,*EKT_Original_II,EKT_Pickup,EKT_Morrison,Morrison_EKT_MCSCF_jcc_1992,IPandEAfromEKT_Cioslowski_jcp_1997,Pernal_IP_EKT_cpl_2005,Dahlenjcp2005,levelsofscGWjcp2009Dahlen,Bozkaya_EKT_jcp_2013}.  It has been a matter of debate whether or not the lowest IP given using EKT is exact, or whether or not higher IPs obtained from this method are physically meaningful\cite{Ernzerhof_Validity_of_EKT_jctc_2009,IPandEAfromEKT_Cioslowski_jcp_1997}. We do not investigate this in this paper, but rather, show that EKT offers reasonable values for IPs and EAs. Presumably, if one were to use a more accurate Green's function, one would obtain more accurate IPs and EAs.

\section{Computational Details}

Our GF2 and G$_{1}$F2 calculations were carried out on an imaginary grid with 20,000 frequency points and 4,400 time points\cite{ALPS}, with an inverse-temperature of $\beta$= 100.0 [$a.u^{-1}$]. Experimentally determined geometries were used for the molecules\cite{geometry}. The basis sets used were Dunning's aug-cc-pVXZ series\cite{basis1,basis2}. Restricted Hartree-Fock calculations carried out in the Dalton program\cite{Dalton} were used to generate an initial Green's function $\mathbf{G}_{0}(\omega)$ as input for our GF2 procedure. All GF2 and G$_1$F2 calculations reported here are all-electron. The Hartree Fock IPs and EAs were obtained from standard Koopmans' Theorem as the negative of the HOMO and LUMO eigenvalues, respectively. To obtain accurate reference theoretical values, IPs and EAs were computed from energy differences of the charged  and  neutral species, using
 all-electron unrestricted UCCSD(T) calculations  with the Gaussian 09 package \cite{g09} (``UCCSD(T)=Full'' keyword). 
Let us stress that since IP and EA values in UCCSD(T) are calculated as a difference between the charged and neutral species, the obtained values not only include the benefit of energy lowering due to the use of an unrestricted method but also can take advantage of error cancellations. This stands in stark contrast to the IPs and EAs calculated from GF2 that is based on a restricted reference (RHF) and does not benefit from error cancellation due to calculating differences.



\onecolumngrid

\section{Results}
\begin{figure}
\includegraphics[width=17cm]{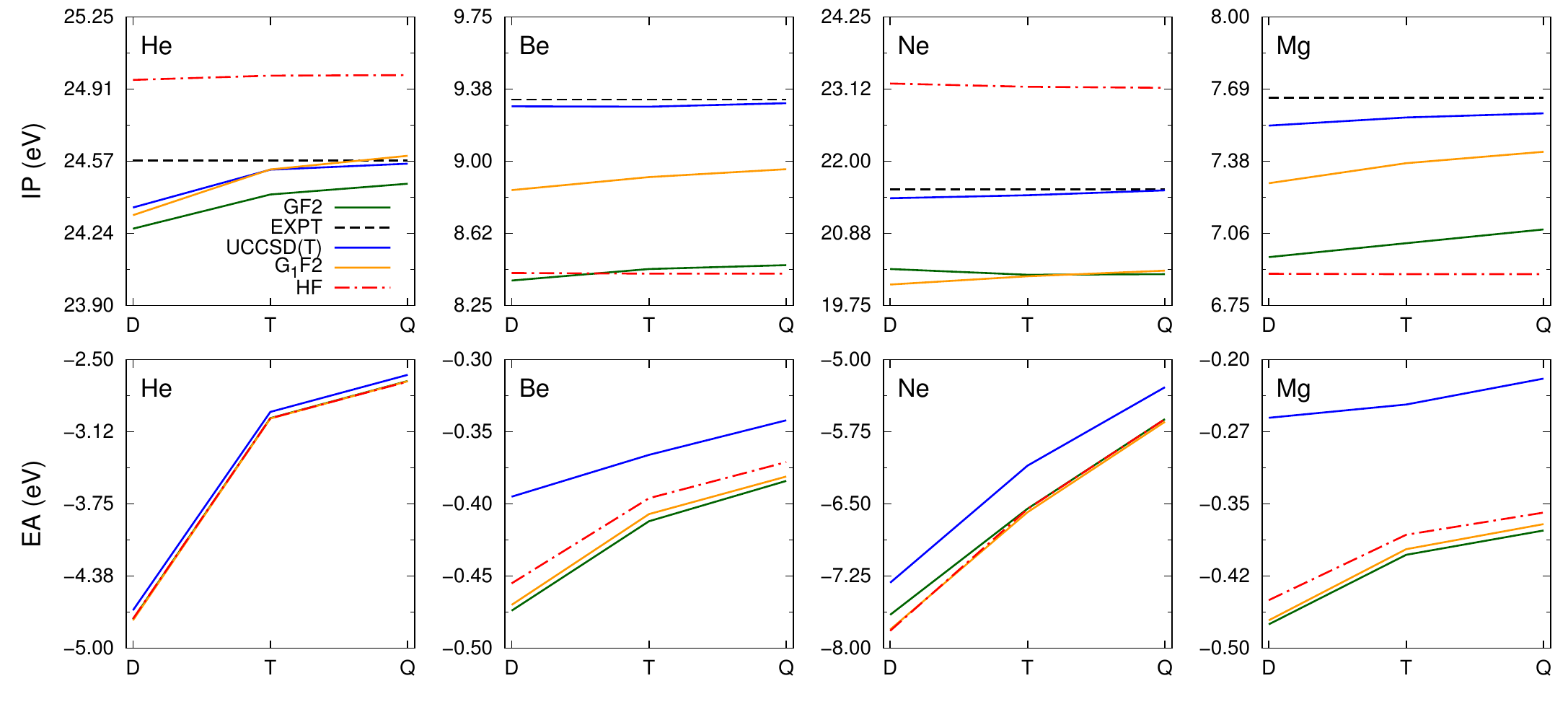}
\caption{Ionization potentials and electron affinities of atoms calculated with the aug-cc-pVXZ basis set.}
\label{fig:IPs_EAs}
\end{figure}

\begin{table}
\setlength{\tabcolsep}{10pt}
\renewcommand{\arraystretch}{1.5}
\begin{tabular}{c | c c   c  c c}
 & GF2 &  G$_1$F2 & HF  &  UCCSD(T) & Expt\\
\hline
Be$_2$&6.21&6.99&6.62&7.42& \\
BH$_3$&12.82&13.13&13.52&13.17& \\
C$_{2}$H$_{2}$&10.24&11.24&11.19&11.36&11.41 \\
C$_{2}$H$_{4}$&9.54&10.22&10.21&10.55&10.51 \\
CO&12.20&14.46&15.09&13.80&14.01 \\
CO$_2$&11.71&12.88&14.82&13.61&13.78 \\
H$_{2}$CO&9.12&9.74&12.02&10.74&10.88 \\
H$_{2}$O&11.31&11.47&13.86&12.54&12.65 \\
H$_2$O$_2$&9.51&10.32&13.31&11.46&11.70 \\
HCN&12.26&13.49&13.50&13.62&13.61 \\
HF&14.68&14.66&17.68&16.02&16.06 \\
Li$_2$&4.69&5.28&4.95&5.23& \\
LiF&9.89&9.70&12.91&11.37& \\
LiH&7.77&7.91&8.20&7.94&7.90 \\
MgH$_2$&9.80&9.93&10.09&9.76& \\
N$_{2}$&13.53&14.97&17.26&15.36&15.58 \\
Na$_2$&4.67&4.80&4.51&4.85& \\
NaF&8.38&8.15&11.59&9.98& \\
NaH&6.82&7.08&7.43&7.04& \\
NaLi&4.69&4.96&4.71&5.01& \\
NaOH&6.42&6.35&9.11&7.86& \\
NH$_3$&9.86&10.10&11.67&10.76&10.82 \\

\end{tabular}

\caption{Calculated ionization potentials in eV  using self-consistent GF2 and non self-consistent G$_1$F2. Shown for comparison are the ionization potentials from Hartree-Fock Koopmans' Theorem and UCCSD(T)  energy-differences, along with experimental values. All calculations use aug-cc-pVDZ.}
\label{tab:IPmol}
\end{table}

\twocolumngrid

Our ability to obtain accurate IPs and EAs will be affected by the intrinsic accuracy of EKT, the performance of GF2, and the choice of basis set. We will not discuss the accuracy of EKT in this work, but will instead focus on the latter two points. To assess the performance of GF2 and G$_1$F2, we have carried out a series of calculations on several closed shell atoms and  molecules. We start from atomic calculations since they are simpler, and then  turn our discussion to small molecules. We have carried out UCCSD(T) calculations on each system to be used as a reference point throughout our discussion. We note that for the closed shell atoms and many of the molecules studied here the EA will be negative, meaning the system does not bind an extra electron, and in the complete basis set limit the EA would approach zero. However we include these systems in our analysis as a proof of concept, because we find the EA from GF2 with EKT for most cases agrees reasonably well with the results from UCCSD(T) energy differences, as well as from HF Koopmans' Theorem.

\subsection{Atoms}

In Figure~\ref{fig:IPs_EAs}, we present for a series of closed-shell atoms the IPs and EAs calculated with GF2 and G$_1$F2 using  EKT, with HF using standard Koopmans' Theorem, and with UCCSD(T) using energy differences, compared against experimental values when available.
Figure~\ref{fig:IPs_EAs} illustrates that the IPs obtained from GF2 for these atoms are well converged with the basis set and systematically underestimate experimental and UCCSD(T) IPs. In comparison with experimental values, the IPs can differ by up to 1 eV. For these atoms the best agreement occurs in the case of He, with a difference from experiment of around 0.1 eV. 
 For comparison, calculations were carried out non self-consistently (G$_{1}$F2), and it was found that these values were in closer agreement with both UCCSD(T) and experimental IPs than the  self-consistent GF2 IPs. 
 This effect was observed previously in the work of Dahlen and van Leeuwen\cite{Dahlenjcp2005,levelsofscGWjcp2009Dahlen}. 
 In contrast, self-consistency appears to have a small effect on the EAs of these systems. The largest difference in electron affinity occurs for Ne, with a difference of 0.2 eV. The majority of the atoms have a difference on the order of 1 meV between self-consistent and non self-consistent calculations. 
It should be noted that the EAs are not converged with respect to the basis set. Between basis sets the EAs can vary by around 1 eV, in a similar fashion to the UCCSD(T) electron affinities. 


\begin{figure}[h]
\includegraphics[width=8.5cm]{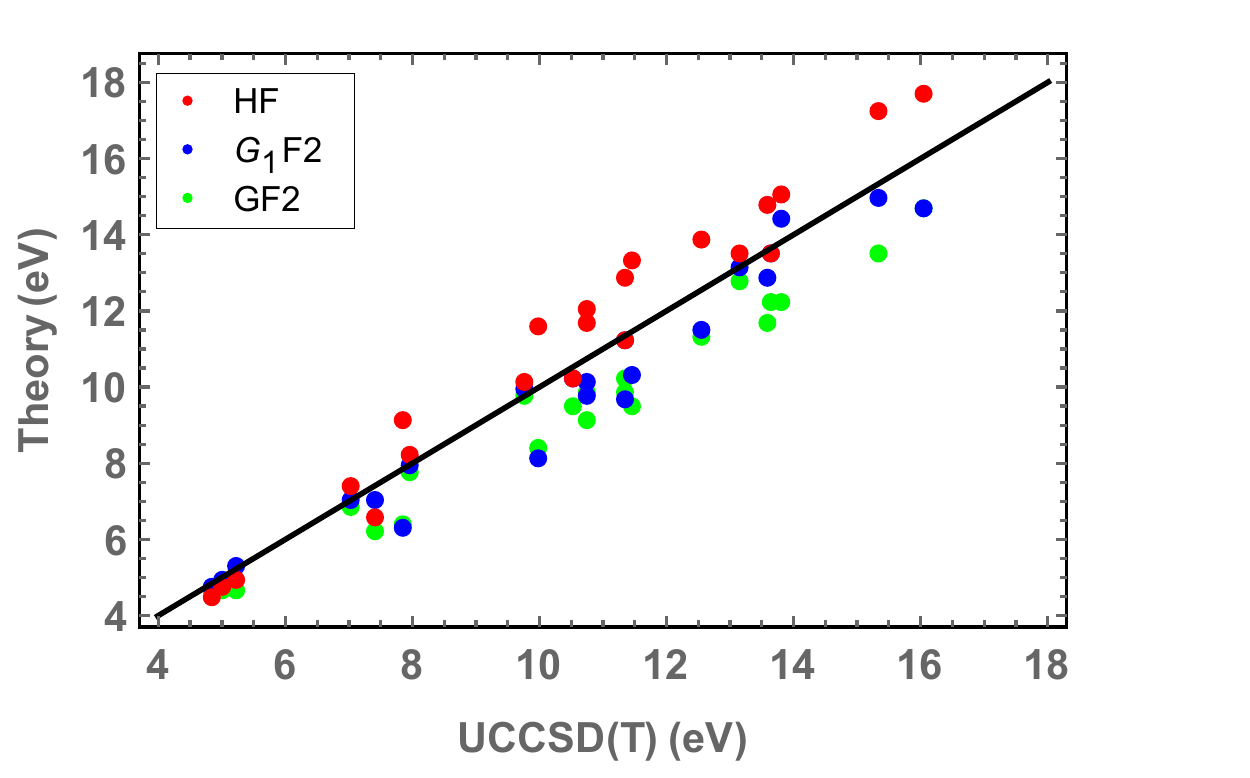}
\caption{HF, G$_1$F2, and GF2 IPs for  molecules compared against reference UCCSD(T) values, calculated with aug-cc-pVDZ.}
\label{fig:IP}
\end{figure}

\subsection{Molecules}
Turning now to the molecules, in Tables~\ref{tab:IPmol} and \ref{tab:EAmol} we present our calculated IPs and EAs. 
 Focusing first on the IPs, similar  to the atoms, we find that GF2 tends to systematically underestimate experimental values.  Furthermore, comparing  against UCCSD(T) reference values in Fig.~\ref{fig:IP} we find that GF2 systematically underestimates UCCSD(T) as well. In contrast, non-self-consistent G$_1$F2 again  yields overall slightly larger IPs that are in better agreement with UCCSD(T) and experimental values than  self-consistent GF2. 
This same effect has been observed in the work of Rostgaard, Jacobsen, and Thygesen\cite{Fully_scGW_Rostgaard_prb_2010} with self-consistent GW and non-self-consistent G$_0$W$_0$ applied to a similar set of molecular systems, where it was interpreted as being caused by overscreening in the former case and underscreening in the latter. GF2 does not have a series of bubble diagrams that are commonly thought to be responsible for the screening effects. However, the self-consistent GF2 has  a series of composite type diagrams resulting from the iterative procedure where four of the diagrams from Figure~\ref{fig:diagrams} are joined together into series of ladder like diagrams. These diagrams  most likely have effects similar to the  series of bubbles in self-consistent GW which are causing overscreening. 
We believe that since the self-consistent GF2 redefines the Fock matrix during  iterations, the results of the iterative procedure are less likely to overestimate the amount of correlation as  in the case of MP2. Therefore, we assume that the good agreement of   G$_1$F2 with UCCSD(T) is fortuitous.
One could speculate that in iterative GF2 introducing the third order diagrams may be much more beneficial and would lead to  systematically convergent IP results.

o

Examining the EAs now, we find  GF2 tends to give results which are slightly lower than both HF and UCCSD(T) values. For a few of the molecules with positive EAs, GF2 and G$_1$F2  do not recover the proper sign. However we emphasize that the EA is a notoriously difficult property to calculate, and from comparing our results to those from EKT-MPn in Table~2 of Ref.~\cite{Bozkaya_EKT_jctc_2014} it is not uncommon for a method to occasionally predict the incorrect sign for even small closed-shell molecules. 
Interestingly, whether or not the calculations are carried out self-consistently does not appear to have such a drastic effect on the EAs, as the GF2, G$_1$F2, and HF EAs tend to be quite similar to the UCCSD(T) values for many of the molecules. We think the simplest explanation for this is the following: in the imaginary time Green's function EKT approach we are using, the EAs are essentially determined by the ``hole''-part of $\mathbf{G}(\tau)$ near $\tau =0_{+}$, and the IPs likewise by the ``particle''-part near $\tau=\beta$. Because these systems are small and weakly correlated, it may be that the ``hole'' or virtual orbital space does not relax as much between HF, G$_{1}$F2, and GF2, as does the ``particle'' or occupied orbital space. Regardless, we find it encouraging that EAs can systematically be recovered from simply the imaginary time Green's function of the neutral system, without  need for considering molecular anions.

\begin{table}
\setlength{\tabcolsep}{10pt}
\renewcommand{\arraystretch}{1.5}
\begin{tabular}{c | c c   c  c }
 & GF2&  G$_1$F2& HF&  UCCSD(T) \\
\hline

Be$_2$ & -0.30& -0.28& -0.21& 0.34 \\
BH$_3$ & -0.92& -0.94& -0.88& -0.15 \\
C$_{2}$H$_{2}$ & -1.02& -1.02& -1.02& -0.99 \\
C$_{2}$H$_{4}$ & -1.09& -1.10& -1.10& -1.22 \\
CO & -2.34& -2.33& -2.15& -1.81 \\
CO$_2$ & -1.46& -1.47& -1.49& -2.23 \\
H$_{2}$CO & -0.94& -0.90& -0.90& -1.19 \\
H$_{2}$O & -0.93& -0.96& -0.96& -0.75 \\
H$_2$O$_2$ & -1.05& -1.08& -1.08& -1.16 \\
HCN & -0.83& -0.79& -0.79& -0.69 \\
HF & -0.95& -0.97& -0.97& -0.80 \\
Li$_2$ & -0.10& -0.10& -0.08& 0.32 \\
LiF & 0.27& 0.29& 0.29& 0.34 \\
LiH & 0.20& 0.20& 0.21& 0.30 \\
MgH$_2$ & -0.45& -0.44& -0.44& -0.50 \\
N$_{2}$ & -2.81 & -4.02& -3.39& -2.60 \\
Na$_2$ & -0.05& -0.05& -0.01& 0.36 \\
NaF & 0.43& 0.43& 0.44& 0.48 \\
NaH & 0.22& 0.23& 0.24& 0.32 \\
NaLi & -0.07& -0.07& -0.04& 0.35 \\
NaOH & 0.32& 0.31& 0.31& 0.39 \\
NH$_3$ & -0.94& -0.97& -0.98& -0.75 \\

\end{tabular}
\caption{Calculated electron affinities in eV  using self-consistent GF2 and non self-consistent G$_1$F2. Shown for comparison are the electron affinities from Hartree-Fock Koopmans' Theorem, and UCCSD(T) energy-differences. All calculations use aug-cc-pVDZ.}
\label{tab:EAmol}
\end{table}




\section{Conclusions}

In this work we have investigated how reliably IPs and EAs can be calculated from the Extended Koopmans' Theorem (EKT) with an imaginary time Green's function in a second order approximation (GF2). In contrast to prior EKT works that determined the EA indirectly as the IP of the anion\cite{Bozkaya_EKT_jctc_2014}, in this work we calculated both IPs and EAs directly from the imaginary time Green's function of the neutral system alone. Overall, we find that self-consistent GF2 with EKT  recovers IPs and EAs that are systematically smaller than UCCSD(T) energy-differences and experiment reference values.  Interestingly, non-self-consistent G$_1$F2 on a Hartree-Fock reference consistently gives slightly larger IPs and EAs than self-consistent GF2, similar to what has been found with GW and G$_0$W$_0$ for IPs\cite{Fully_scGW_Rostgaard_prb_2010}. Because GF2 is defined in terms of the bare Coulomb interaction rather than the screened interaction as in GW, this suggests that the cause of the systematic underestimation of quasiparticle spectra by self-consistent vs non-self-consistent Green's function methods may be more general than being specifically the result of over or underscreening caused by a series of bubble diagrams present in the the GW approach. 

Regardless of the particular performance of GF2 or G$_1$F2, the more general point of this work is that the EKT in conjunction with self-consistent Green's function theory offers a reliable procedure for the unbiased theoretical determination of quasiparticle spectra.
 Essentially the underlying scheme in this Green's function EKT approach is that the IPs are determined by the eigenvalues of the time-derivative of the ``particle''-part of $\mathbf{G}(\tau)$ at $\tau=\beta$, while the EAs are likewise found from the ``hole''-part of $\mathbf{G}(\tau)$ at $\tau=0_+$. In this way the full quasiparticle spectra can in principle be reconstructed from simply 
 the Green's function on the imaginary time domain, without the need for analytic continuation\cite{Analytic_cont1, Analytic_cont2, Analytic_cont3} or other numerical methods.
 Therefore the EKT allows one to obtain real axis quasiparticle spectra while still enjoying the computational benefits of using an imaginary axis Green's function implementation.
 Furthermore, the advantages of the EKT are that it is simple, is applicable to a Green's function from any level of theory, requires only a trivial amount of computational time, and can be implemented in a blackbox manner.


\section{Acknowledgments }
D.  Zgid, A. Welden and  J.J.  Phillips  acknowledge  support  from
a  DOE  grant  no.   ER16391  and  an  XSEDE  allocation
allowing us to use the STAMPEDE supercomputer.



\providecommand*\mcitethebibliography{\thebibliography}
\csname @ifundefined\endcsname{endmcitethebibliography}
  {\let\endmcitethebibliography\endthebibliography}{}


\begin{mcitethebibliography}{129}
\providecommand*\natexlab[1]{#1}
\providecommand*\mciteSetBstSublistMode[1]{}
\providecommand*\mciteSetBstMaxWidthForm[2]{}
\providecommand*\mciteBstWouldAddEndPuncttrue
  {\def\EndOfBibitem{\unskip.}}
\providecommand*\mciteBstWouldAddEndPunctfalse
  {\let\EndOfBibitem\relax}
\providecommand*\mciteSetBstMidEndSepPunct[3]{}
\providecommand*\mciteSetBstSublistLabelBeginEnd[3]{}
\providecommand*\EndOfBibitem{}
\mciteSetBstSublistMode{f}
\mciteSetBstMaxWidthForm{subitem}{(\alph{mcitesubitemcount})}
\mciteSetBstSublistLabelBeginEnd
  {\mcitemaxwidthsubitemform\space}
  {\relax}
  {\relax}

\bibitem[Hohenberg and Kohn(1964)Hohenberg, and Kohn]{PhysRev.136.B864}
Hohenberg,~P.; Kohn,~W. \emph{Phys. Rev.} \textbf{1964}, \emph{136},
  B864--B871\relax
\mciteBstWouldAddEndPuncttrue
\mciteSetBstMidEndSepPunct{\mcitedefaultmidpunct}
{\mcitedefaultendpunct}{\mcitedefaultseppunct}\relax
\EndOfBibitem
\bibitem[Kohn and Sham(1965)Kohn, and Sham]{PhysRev.140.A1133}
Kohn,~W.; Sham,~L.~J. \emph{Phys. Rev.} \textbf{1965}, \emph{140},
  A1133--A1138\relax
\mciteBstWouldAddEndPuncttrue
\mciteSetBstMidEndSepPunct{\mcitedefaultmidpunct}
{\mcitedefaultendpunct}{\mcitedefaultseppunct}\relax
\EndOfBibitem
\bibitem[Parr and Yang(1989)Parr, and Yang]{DFTbookParr}
Parr,~R.~G.; Yang,~W. \emph{Density-Functional Theory of Atoms and Molecules};
  Oxford University Press: New York, 1989\relax
\mciteBstWouldAddEndPuncttrue
\mciteSetBstMidEndSepPunct{\mcitedefaultmidpunct}
{\mcitedefaultendpunct}{\mcitedefaultseppunct}\relax
\EndOfBibitem
\bibitem[Coester and K\"ummel(1960)Coester, and
  K\"ummel]{CoupledClusterTheory_nucphys_1960}
Coester,~F.; K\"ummel,~H. \emph{Nuclear Physics} \textbf{1960}, \emph{17}, 477
  -- 485\relax
\mciteBstWouldAddEndPuncttrue
\mciteSetBstMidEndSepPunct{\mcitedefaultmidpunct}
{\mcitedefaultendpunct}{\mcitedefaultseppunct}\relax
\EndOfBibitem
\bibitem[Cizek(1991)]{CC_Cizek}
Cizek,~J. \emph{Theoretica chimica acta} \textbf{1991}, \emph{80}, 91--94\relax
\mciteBstWouldAddEndPuncttrue
\mciteSetBstMidEndSepPunct{\mcitedefaultmidpunct}
{\mcitedefaultendpunct}{\mcitedefaultseppunct}\relax
\EndOfBibitem
\bibitem[Bartlett and Musia\l{}(2007)Bartlett, and Musia\l{}]{CCBartlett07}
Bartlett,~R.~J.; Musia\l{},~M. \emph{Rev. Mod. Phys.} \textbf{2007}, \emph{79},
  291--352\relax
\mciteBstWouldAddEndPuncttrue
\mciteSetBstMidEndSepPunct{\mcitedefaultmidpunct}
{\mcitedefaultendpunct}{\mcitedefaultseppunct}\relax
\EndOfBibitem
\bibitem[Scuseria(1999)]{ScusScale1999}
Scuseria,~G.~E. \emph{J. Phys. Chem. A} \textbf{1999}, \emph{103},
  4782--4790\relax
\mciteBstWouldAddEndPuncttrue
\mciteSetBstMidEndSepPunct{\mcitedefaultmidpunct}
{\mcitedefaultendpunct}{\mcitedefaultseppunct}\relax
\EndOfBibitem
\bibitem[Burke(2012)]{Burke_Perspective_jcp_2012}
Burke,~K. \emph{J. Chem. Phys.} \textbf{2012}, \emph{136}, --\relax
\mciteBstWouldAddEndPuncttrue
\mciteSetBstMidEndSepPunct{\mcitedefaultmidpunct}
{\mcitedefaultendpunct}{\mcitedefaultseppunct}\relax
\EndOfBibitem
\bibitem[Bredow and Gerson(2000)Bredow, and Gerson]{EffectofXC_Gerson_prb_2000}
Bredow,~T.; Gerson,~A.~R. \emph{Phys. Rev. B} \textbf{2000}, \emph{61},
  5194--5201\relax
\mciteBstWouldAddEndPuncttrue
\mciteSetBstMidEndSepPunct{\mcitedefaultmidpunct}
{\mcitedefaultendpunct}{\mcitedefaultseppunct}\relax
\EndOfBibitem
\bibitem[de~P.~R.~Moreira et~al.(2002)de~P.~R.~Moreira, Illas, and
  Martin]{Moreira:2002ad}
de~P.~R.~Moreira,~I.; Illas,~F.; Martin,~R.~L. \emph{Phys. Rev. B}
  \textbf{2002}, \emph{65}, 155102\relax
\mciteBstWouldAddEndPuncttrue
\mciteSetBstMidEndSepPunct{\mcitedefaultmidpunct}
{\mcitedefaultendpunct}{\mcitedefaultseppunct}\relax
\EndOfBibitem
\bibitem[Perdew and Levy(1983)Perdew, and Levy]{Perdew_KS_Orbitals_1983}
Perdew,~J.~P.; Levy,~M. \emph{Phys. Rev. Lett.} \textbf{1983}, \emph{51},
  1884--1887\relax
\mciteBstWouldAddEndPuncttrue
\mciteSetBstMidEndSepPunct{\mcitedefaultmidpunct}
{\mcitedefaultendpunct}{\mcitedefaultseppunct}\relax
\EndOfBibitem
\bibitem[Perdew(1985)]{Perdew_BandGapProblem_ijqc_1985}
Perdew,~J.~P. \emph{Int. J. Quantum Chem.} \textbf{1985}, \emph{28},
  497--523\relax
\mciteBstWouldAddEndPuncttrue
\mciteSetBstMidEndSepPunct{\mcitedefaultmidpunct}
{\mcitedefaultendpunct}{\mcitedefaultseppunct}\relax
\EndOfBibitem
\bibitem[Stoudenmire et~al.(2012)Stoudenmire, Wagner, White, and
  Burke]{OneDDMRGDFT_Burke_prl_2012}
Stoudenmire,~E.~M.; Wagner,~L.~O.; White,~S.~R.; Burke,~K. \emph{Phys. Rev.
  Lett.} \textbf{2012}, \emph{109}, 056402\relax
\mciteBstWouldAddEndPuncttrue
\mciteSetBstMidEndSepPunct{\mcitedefaultmidpunct}
{\mcitedefaultendpunct}{\mcitedefaultseppunct}\relax
\EndOfBibitem
\bibitem[Baerends et~al.(2013)Baerends, Gritsenko, and van
  Meer]{KSgap_Fundgap_Optgap_PCCP_2013}
Baerends,~E.~J.; Gritsenko,~O.~V.; van Meer,~R. \emph{Phys. Chem. Chem. Phys.}
  \textbf{2013}, \emph{15}, 16408--16425\relax
\mciteBstWouldAddEndPuncttrue
\mciteSetBstMidEndSepPunct{\mcitedefaultmidpunct}
{\mcitedefaultendpunct}{\mcitedefaultseppunct}\relax
\EndOfBibitem
\bibitem[Foerster et~al.(2011)Foerster, Koval, and
  Sánchez-Portal]{GoWo_Koval_jcp_2011}
Foerster,~D.; Koval,~P.; Sánchez-Portal,~D. \emph{J. Chem. Phys.}
  \textbf{2011}, \emph{135}, --\relax
\mciteBstWouldAddEndPuncttrue
\mciteSetBstMidEndSepPunct{\mcitedefaultmidpunct}
{\mcitedefaultendpunct}{\mcitedefaultseppunct}\relax
\EndOfBibitem
\bibitem[Caruso et~al.(2013)Caruso, Rinke, Ren, Rubio, and
  Scheffler]{scGWprb2013Scheffler}
Caruso,~F.; Rinke,~P.; Ren,~X.; Rubio,~A.; Scheffler,~M. \emph{Phys. Rev. B}
  \textbf{2013}, \emph{88}, 075105\relax
\mciteBstWouldAddEndPuncttrue
\mciteSetBstMidEndSepPunct{\mcitedefaultmidpunct}
{\mcitedefaultendpunct}{\mcitedefaultseppunct}\relax
\EndOfBibitem
\bibitem[Phillips and Zgid(2014)Phillips, and Zgid]{gf2_paper_2014}
Phillips,~J.~J.; Zgid,~D. \emph{J. Chem. Phys.} \textbf{2014}, \emph{140},
  --\relax
\mciteBstWouldAddEndPuncttrue
\mciteSetBstMidEndSepPunct{\mcitedefaultmidpunct}
{\mcitedefaultendpunct}{\mcitedefaultseppunct}\relax
\EndOfBibitem
\bibitem[Koval et~al.(2014)Koval, Foerster, and
  S\'anchez-Portal]{FullySCGW_Koval_prb_2014}
Koval,~P.; Foerster,~D.; S\'anchez-Portal,~D. \emph{Phys. Rev. B}
  \textbf{2014}, \emph{89}, 155417\relax
\mciteBstWouldAddEndPuncttrue
\mciteSetBstMidEndSepPunct{\mcitedefaultmidpunct}
{\mcitedefaultendpunct}{\mcitedefaultseppunct}\relax
\EndOfBibitem
\bibitem[Dahlen and van Leeuwen(2005)Dahlen, and van Leeuwen]{Dahlenjcp2005}
Dahlen,~N.~E.; van Leeuwen,~R. \emph{J. Chem. Phys.} \textbf{2005}, \emph{122},
  --\relax
\mciteBstWouldAddEndPuncttrue
\mciteSetBstMidEndSepPunct{\mcitedefaultmidpunct}
{\mcitedefaultendpunct}{\mcitedefaultseppunct}\relax
\EndOfBibitem
\bibitem[Stan et~al.(2009)Stan, Dahlen, and van
  Leeuwen]{levelsofscGWjcp2009Dahlen}
Stan,~A.; Dahlen,~N.~E.; van Leeuwen,~R. \emph{J. Chem. Phys.} \textbf{2009},
  \emph{130}, --\relax
\mciteBstWouldAddEndPuncttrue
\mciteSetBstMidEndSepPunct{\mcitedefaultmidpunct}
{\mcitedefaultendpunct}{\mcitedefaultseppunct}\relax
\EndOfBibitem
\bibitem[M\o{}ller and Plesset(1934)M\o{}ller, and Plesset]{MollerPlesset1934}
M\o{}ller,~C.; Plesset,~M.~S. \emph{Phys. Rev.} \textbf{1934}, \emph{46},
  618--622\relax
\mciteBstWouldAddEndPuncttrue
\mciteSetBstMidEndSepPunct{\mcitedefaultmidpunct}
{\mcitedefaultendpunct}{\mcitedefaultseppunct}\relax
\EndOfBibitem
\bibitem[Hedin(1965)]{HedinGWpra1965}
Hedin,~L. \emph{Phys. Rev.} \textbf{1965}, \emph{139}, A796--A823\relax
\mciteBstWouldAddEndPuncttrue
\mciteSetBstMidEndSepPunct{\mcitedefaultmidpunct}
{\mcitedefaultendpunct}{\mcitedefaultseppunct}\relax
\EndOfBibitem
\bibitem[Albertsen and J\o{}rgensen(1979)Albertsen, and
  J\o{}rgensen]{IPsEAs_2ndOrderEPT_Jorgensen_jcp_1979}
Albertsen,~P.; J\o{}rgensen,~P. \emph{J. Chem. Phys.} \textbf{1979}, \emph{70},
  3254--3263\relax
\mciteBstWouldAddEndPuncttrue
\mciteSetBstMidEndSepPunct{\mcitedefaultmidpunct}
{\mcitedefaultendpunct}{\mcitedefaultseppunct}\relax
\EndOfBibitem
\bibitem[B. and Öhrn(1981)B., and Öhrn]{Ortiz_EPTofEAs_cpl_1981}
B.,~J.~O.; Öhrn,~Y. \emph{Chem. Phys. Lett.} \textbf{1981}, \emph{77}, 548 --
  554\relax
\mciteBstWouldAddEndPuncttrue
\mciteSetBstMidEndSepPunct{\mcitedefaultmidpunct}
{\mcitedefaultendpunct}{\mcitedefaultseppunct}\relax
\EndOfBibitem
\bibitem[Schirmer et~al.(1983)Schirmer, Cederbaum, and Walter]{ADC_pra_1983}
Schirmer,~J.; Cederbaum,~L.~S.; Walter,~O. \emph{Phys. Rev. A} \textbf{1983},
  \emph{28}, 1237--1259\relax
\mciteBstWouldAddEndPuncttrue
\mciteSetBstMidEndSepPunct{\mcitedefaultmidpunct}
{\mcitedefaultendpunct}{\mcitedefaultseppunct}\relax
\EndOfBibitem
\bibitem[Cederbaum(1990)]{OnGfAndTheirApplicationsCederbaum1990}
Cederbaum,~L.~S. \emph{Int. J. Quantum Chem.} \textbf{1990}, \emph{38},
  393--404\relax
\mciteBstWouldAddEndPuncttrue
\mciteSetBstMidEndSepPunct{\mcitedefaultmidpunct}
{\mcitedefaultendpunct}{\mcitedefaultseppunct}\relax
\EndOfBibitem
\bibitem[Massidda et~al.(1995)Massidda, Continenza, Posternak, and
  Baldereschi]{Massidda_MnO_GW_prl_1995}
Massidda,~S.; Continenza,~A.; Posternak,~M.; Baldereschi,~A. \emph{Phys. Rev.
  Lett.} \textbf{1995}, \emph{74}, 2323--2326\relax
\mciteBstWouldAddEndPuncttrue
\mciteSetBstMidEndSepPunct{\mcitedefaultmidpunct}
{\mcitedefaultendpunct}{\mcitedefaultseppunct}\relax
\EndOfBibitem
\bibitem[Massidda et~al.(1997)Massidda, Continenza, Posternak, and
  Baldereschi]{Massidda_MetalOxide_GW_prb_1997}
Massidda,~S.; Continenza,~A.; Posternak,~M.; Baldereschi,~A. \emph{Phys. Rev.
  B} \textbf{1997}, \emph{55}, 13494--13502\relax
\mciteBstWouldAddEndPuncttrue
\mciteSetBstMidEndSepPunct{\mcitedefaultmidpunct}
{\mcitedefaultendpunct}{\mcitedefaultseppunct}\relax
\EndOfBibitem
\bibitem[Ku and Eguiluz(2002)Ku, and
  Eguiluz]{BandGapProblem_scGW_KuWei_prl_2002}
Ku,~W.; Eguiluz,~A.~G. \emph{Phys. Rev. Lett.} \textbf{2002}, \emph{89},
  126401\relax
\mciteBstWouldAddEndPuncttrue
\mciteSetBstMidEndSepPunct{\mcitedefaultmidpunct}
{\mcitedefaultendpunct}{\mcitedefaultseppunct}\relax
\EndOfBibitem
\bibitem[Rostgaard et~al.(2010)Rostgaard, Jacobsen, and
  Thygesen]{Fully_scGW_Rostgaard_prb_2010}
Rostgaard,~C.; Jacobsen,~K.~W.; Thygesen,~K.~S. \emph{Phys. Rev. B}
  \textbf{2010}, \emph{81}, 085103\relax
\mciteBstWouldAddEndPuncttrue
\mciteSetBstMidEndSepPunct{\mcitedefaultmidpunct}
{\mcitedefaultendpunct}{\mcitedefaultseppunct}\relax
\EndOfBibitem
\bibitem[Ortiz(2013)]{Ortiz_Review_2013}
Ortiz,~J.~V. \emph{Wiley Interdisciplinary Reviews: Computational Molecular
  Science} \textbf{2013}, \emph{3}, 123--142\relax
\mciteBstWouldAddEndPuncttrue
\mciteSetBstMidEndSepPunct{\mcitedefaultmidpunct}
{\mcitedefaultendpunct}{\mcitedefaultseppunct}\relax
\EndOfBibitem
\bibitem[Gunnarsson et~al.(2010)Gunnarsson, Haverkort, and
  Sangiovanni]{Analytic_cont1}
Gunnarsson,~O.; Haverkort,~M.~W.; Sangiovanni,~G. \emph{Phys. Rev. B}
  \textbf{2010}, \emph{82}, 165125\relax
\mciteBstWouldAddEndPuncttrue
\mciteSetBstMidEndSepPunct{\mcitedefaultmidpunct}
{\mcitedefaultendpunct}{\mcitedefaultseppunct}\relax
\EndOfBibitem
\bibitem[Jarrell and Gubernatis(1996)Jarrell, and Gubernatis]{Analytic_cont2}
Jarrell,~M.; Gubernatis,~J.~E. \emph{Physics Reports} \textbf{1996},
  \emph{269}, 133--195\relax
\mciteBstWouldAddEndPuncttrue
\mciteSetBstMidEndSepPunct{\mcitedefaultmidpunct}
{\mcitedefaultendpunct}{\mcitedefaultseppunct}\relax
\EndOfBibitem
\bibitem[Gubernatis et~al.(1991)Gubernatis, Jarrell, Silver, and
  Sivia]{Analytic_cont3}
Gubernatis,~J.; Jarrell,~M.; Silver,~R.; Sivia,~D. \emph{Physical Review B}
  \textbf{1991}, \emph{44}, 6011\relax
\mciteBstWouldAddEndPuncttrue
\mciteSetBstMidEndSepPunct{\mcitedefaultmidpunct}
{\mcitedefaultendpunct}{\mcitedefaultseppunct}\relax
\EndOfBibitem
\bibitem[Smith and Day(1975)Smith, and Day]{EKT_Original_jcp_1975}
Smith,~D.~W.; Day,~O.~W. \emph{The Journal of Chemical Physics} \textbf{1975},
  \emph{62}, 113--114\relax
\mciteBstWouldAddEndPuncttrue
\mciteSetBstMidEndSepPunct{\mcitedefaultmidpunct}
{\mcitedefaultendpunct}{\mcitedefaultseppunct}\relax
\EndOfBibitem
\bibitem[Day et~al.(1975)Day, Smith, and Morrison]{EKT_Original_II}
Day,~O.~W.; Smith,~D.~W.; Morrison,~R.~C. \emph{The Journal of Chemical
  Physics} \textbf{1975}, \emph{62}, 115--119\relax
\mciteBstWouldAddEndPuncttrue
\mciteSetBstMidEndSepPunct{\mcitedefaultmidpunct}
{\mcitedefaultendpunct}{\mcitedefaultseppunct}\relax
\EndOfBibitem
\bibitem[Pickup(1975)]{EKT_Pickup}
Pickup,~B.~T. \emph{Chem. Phys. Lett.} \textbf{1975}, \emph{33}, 422 --
  426\relax
\mciteBstWouldAddEndPuncttrue
\mciteSetBstMidEndSepPunct{\mcitedefaultmidpunct}
{\mcitedefaultendpunct}{\mcitedefaultseppunct}\relax
\EndOfBibitem
\bibitem[Katriel and Davidson(1980)Katriel, and
  Davidson]{KatrielDavidson_EKT_pnas_1980}
Katriel,~J.; Davidson,~E.~R. \emph{Proceedings of the National Academy of
  Sciences} \textbf{1980}, \emph{77}, 4403--4406\relax
\mciteBstWouldAddEndPuncttrue
\mciteSetBstMidEndSepPunct{\mcitedefaultmidpunct}
{\mcitedefaultendpunct}{\mcitedefaultseppunct}\relax
\EndOfBibitem
\bibitem[Matos and Day(1987)Matos, and Day]{EKT_IJQC}
Matos,~J. M.~O.; Day,~O.~W. \emph{International Journal of Quantum Chemistry}
  \textbf{1987}, \emph{31}, 871--892\relax
\mciteBstWouldAddEndPuncttrue
\mciteSetBstMidEndSepPunct{\mcitedefaultmidpunct}
{\mcitedefaultendpunct}{\mcitedefaultseppunct}\relax
\EndOfBibitem
\bibitem[Morrison(1992)]{EKT_Morrison}
Morrison,~R.~C. \emph{J. Chem. Phys.} \textbf{1992}, \emph{96},
  3718--3722\relax
\mciteBstWouldAddEndPuncttrue
\mciteSetBstMidEndSepPunct{\mcitedefaultmidpunct}
{\mcitedefaultendpunct}{\mcitedefaultseppunct}\relax
\EndOfBibitem
\bibitem[Sundholm and Olsen(1993)Sundholm, and Olsen]{Olsen_EKT_jcp_1993}
Sundholm,~D.; Olsen,~J. \emph{The Journal of Chemical Physics} \textbf{1993},
  \emph{98}, 3999--4002\relax
\mciteBstWouldAddEndPuncttrue
\mciteSetBstMidEndSepPunct{\mcitedefaultmidpunct}
{\mcitedefaultendpunct}{\mcitedefaultseppunct}\relax
\EndOfBibitem
\bibitem[Morrison and Ayers(1995)Morrison, and Ayers]{AyersEKT_jcp_1995}
Morrison,~R.~C.; Ayers,~P.~W. \emph{J. Chem. Phys.} \textbf{1995}, \emph{103},
  6556--6561\relax
\mciteBstWouldAddEndPuncttrue
\mciteSetBstMidEndSepPunct{\mcitedefaultmidpunct}
{\mcitedefaultendpunct}{\mcitedefaultseppunct}\relax
\EndOfBibitem
\bibitem[Cioslowski et~al.(1997)Cioslowski, Piskorz, and
  Liu]{IPandEAfromEKT_Cioslowski_jcp_1997}
Cioslowski,~J.; Piskorz,~P.; Liu,~G. \emph{The Journal of Chemical Physics}
  \textbf{1997}, \emph{107}, 6804--6811\relax
\mciteBstWouldAddEndPuncttrue
\mciteSetBstMidEndSepPunct{\mcitedefaultmidpunct}
{\mcitedefaultendpunct}{\mcitedefaultseppunct}\relax
\EndOfBibitem
\bibitem[Olsen and Sundholm(1998)Olsen, and
  Sundholm]{Olsen_PertExpansion_EKT_cpl_1998}
Olsen,~J.; Sundholm,~D. \emph{cpl} \textbf{1998}, \emph{288}, 282 -- 288\relax
\mciteBstWouldAddEndPuncttrue
\mciteSetBstMidEndSepPunct{\mcitedefaultmidpunct}
{\mcitedefaultendpunct}{\mcitedefaultseppunct}\relax
\EndOfBibitem
\bibitem[Pernal and Cioslowski(2001)Pernal, and Cioslowski]{EKT_Cioslowski2001}
Pernal,~K.; Cioslowski,~J. \emph{J. Chem. Phys.} \textbf{2001}, \emph{114},
  4359--4361\relax
\mciteBstWouldAddEndPuncttrue
\mciteSetBstMidEndSepPunct{\mcitedefaultmidpunct}
{\mcitedefaultendpunct}{\mcitedefaultseppunct}\relax
\EndOfBibitem
\bibitem[Ernzerhof(2009)]{Ernzerhof_Validity_of_EKT_jctc_2009}
Ernzerhof,~M. \emph{J. Chem. Theory Comput.} \textbf{2009}, \emph{5},
  793--797\relax
\mciteBstWouldAddEndPuncttrue
\mciteSetBstMidEndSepPunct{\mcitedefaultmidpunct}
{\mcitedefaultendpunct}{\mcitedefaultseppunct}\relax
\EndOfBibitem
\bibitem[Bozkaya(2013)]{Bozkaya_EKT_jcp_2013}
Bozkaya,~U. \emph{The Journal of Chemical Physics} \textbf{2013}, \emph{139},
  --\relax
\mciteBstWouldAddEndPuncttrue
\mciteSetBstMidEndSepPunct{\mcitedefaultmidpunct}
{\mcitedefaultendpunct}{\mcitedefaultseppunct}\relax
\EndOfBibitem
\bibitem[Bozkaya(2014)]{Bozkaya_EKT_jctc_2014}
Bozkaya,~U. \emph{Journal of Chemical Theory and Computation} \textbf{2014},
  \emph{10}, 2041--2048\relax
\mciteBstWouldAddEndPuncttrue
\mciteSetBstMidEndSepPunct{\mcitedefaultmidpunct}
{\mcitedefaultendpunct}{\mcitedefaultseppunct}\relax
\EndOfBibitem
\bibitem[Reed et~al.(1985)Reed, Weinstock, and Weinhold]{Reed_NPA_jcp_1985}
Reed,~A.~E.; Weinstock,~R.~B.; Weinhold,~F. \emph{J. Chem. Phys.}
  \textbf{1985}, \emph{83}, 735\relax
\mciteBstWouldAddEndPuncttrue
\mciteSetBstMidEndSepPunct{\mcitedefaultmidpunct}
{\mcitedefaultendpunct}{\mcitedefaultseppunct}\relax
\EndOfBibitem
\bibitem[Morrison and Liu(1992)Morrison, and Liu]{Morrison_EKT_MCSCF_jcc_1992}
Morrison,~R.~C.; Liu,~G. \emph{J. Comp. Chem.} \textbf{1992}, \emph{13},
  1004--1010\relax
\mciteBstWouldAddEndPuncttrue
\mciteSetBstMidEndSepPunct{\mcitedefaultmidpunct}
{\mcitedefaultendpunct}{\mcitedefaultseppunct}\relax
\EndOfBibitem
\bibitem[Pernal and Cioslowski(2005)Pernal, and
  Cioslowski]{Pernal_IP_EKT_cpl_2005}
Pernal,~K.; Cioslowski,~J. \emph{Chem. Phys. Lett.} \textbf{2005}, \emph{412},
  71 -- 75\relax
\mciteBstWouldAddEndPuncttrue
\mciteSetBstMidEndSepPunct{\mcitedefaultmidpunct}
{\mcitedefaultendpunct}{\mcitedefaultseppunct}\relax
\EndOfBibitem
\bibitem[Sasaki and Yoshimine(1974)Sasaki, and Yoshimine]{Sasaki_EA_PRA_1974_i}
Sasaki,~F.; Yoshimine,~M. \emph{Phys. Rev. A} \textbf{1974}, \emph{9},
  17--25\relax
\mciteBstWouldAddEndPuncttrue
\mciteSetBstMidEndSepPunct{\mcitedefaultmidpunct}
{\mcitedefaultendpunct}{\mcitedefaultseppunct}\relax
\EndOfBibitem
\bibitem[Sasaki and Yoshimine(1974)Sasaki, and
  Yoshimine]{Sasaki_EA_PRA_1974_ii}
Sasaki,~F.; Yoshimine,~M. \emph{Phys. Rev. A} \textbf{1974}, \emph{9},
  26--34\relax
\mciteBstWouldAddEndPuncttrue
\mciteSetBstMidEndSepPunct{\mcitedefaultmidpunct}
{\mcitedefaultendpunct}{\mcitedefaultseppunct}\relax
\EndOfBibitem
\bibitem[Feller and Davidson(1989)Feller, and
  Davidson]{Feller_Davison_EAofOxy_jcp_1989}
Feller,~D.; Davidson,~E.~R. \emph{J. Chem. Phys.} \textbf{1989}, \emph{90},
  1024--1030\relax
\mciteBstWouldAddEndPuncttrue
\mciteSetBstMidEndSepPunct{\mcitedefaultmidpunct}
{\mcitedefaultendpunct}{\mcitedefaultseppunct}\relax
\EndOfBibitem
\bibitem[Kendall et~al.(1992)Kendall, Dunning, and
  Harrison]{Dunning_ElecAffinity_jcp_1992}
Kendall,~R.~A.; Dunning,~T.~H.; Harrison,~R.~J. \emph{J. Chem. Phys.}
  \textbf{1992}, \emph{96}, 6796--6806\relax
\mciteBstWouldAddEndPuncttrue
\mciteSetBstMidEndSepPunct{\mcitedefaultmidpunct}
{\mcitedefaultendpunct}{\mcitedefaultseppunct}\relax
\EndOfBibitem
\bibitem[R\"{o}sch and Trickey(1997)R\"{o}sch, and Trickey]{CommentAnion1996}
R\"{o}sch,~N.; Trickey,~S.~B. \emph{J. Chem. Phys.} \textbf{1997}, \emph{106},
  8940--8941\relax
\mciteBstWouldAddEndPuncttrue
\mciteSetBstMidEndSepPunct{\mcitedefaultmidpunct}
{\mcitedefaultendpunct}{\mcitedefaultseppunct}\relax
\EndOfBibitem
\bibitem[Galbraith and Schaefer(1996)Galbraith, and
  Schaefer]{ConcerningAnion1996}
Galbraith,~J.~M.; Schaefer,~H.~F. \emph{J. Chem. Phys.} \textbf{1996},
  \emph{105}, 862--864\relax
\mciteBstWouldAddEndPuncttrue
\mciteSetBstMidEndSepPunct{\mcitedefaultmidpunct}
{\mcitedefaultendpunct}{\mcitedefaultseppunct}\relax
\EndOfBibitem
\bibitem[Jensen(2010)]{JensenAnion2010}
Jensen,~F. \emph{J. Chem. Theory Comput.} \textbf{2010}, \emph{6},
  2726--2735\relax
\mciteBstWouldAddEndPuncttrue
\mciteSetBstMidEndSepPunct{\mcitedefaultmidpunct}
{\mcitedefaultendpunct}{\mcitedefaultseppunct}\relax
\EndOfBibitem
\bibitem[Emrich(1981)]{EOMCC_Emrich_nucphysa_1981}
Emrich,~K. \emph{Nuclear Physics A} \textbf{1981}, \emph{351}, 379 -- 396\relax
\mciteBstWouldAddEndPuncttrue
\mciteSetBstMidEndSepPunct{\mcitedefaultmidpunct}
{\mcitedefaultendpunct}{\mcitedefaultseppunct}\relax
\EndOfBibitem
\bibitem[Sekino and Bartlett(1984)Sekino, and Bartlett]{EOMCC_Sekino_ijqc_1984}
Sekino,~H.; Bartlett,~R.~J. \emph{International Journal of Quantum Chemistry}
  \textbf{1984}, \emph{26}, 255--265\relax
\mciteBstWouldAddEndPuncttrue
\mciteSetBstMidEndSepPunct{\mcitedefaultmidpunct}
{\mcitedefaultendpunct}{\mcitedefaultseppunct}\relax
\EndOfBibitem
\bibitem[Geertsen et~al.(1989)Geertsen, Rittby, and
  Bartlett]{EOMCC_Geertsen_cpl_1989}
Geertsen,~J.; Rittby,~M.; Bartlett,~R.~J. \emph{Chem. Phys. Lett.}
  \textbf{1989}, \emph{164}, 57 -- 62\relax
\mciteBstWouldAddEndPuncttrue
\mciteSetBstMidEndSepPunct{\mcitedefaultmidpunct}
{\mcitedefaultendpunct}{\mcitedefaultseppunct}\relax
\EndOfBibitem
\bibitem[Stanton and Bartlett(1993)Stanton, and
  Bartlett]{EOMCC_Stanton_jcp_1993}
Stanton,~J.~F.; Bartlett,~R.~J. \emph{J. Chem. Phys.} \textbf{1993}, \emph{98},
  7029--7039\relax
\mciteBstWouldAddEndPuncttrue
\mciteSetBstMidEndSepPunct{\mcitedefaultmidpunct}
{\mcitedefaultendpunct}{\mcitedefaultseppunct}\relax
\EndOfBibitem
\bibitem[Krylov(2008)]{EOM_Krylov_arpc_2008}
Krylov,~A.~I. \emph{Annual Review of Physical Chemistry} \textbf{2008},
  \emph{59}, 433--462\relax
\mciteBstWouldAddEndPuncttrue
\mciteSetBstMidEndSepPunct{\mcitedefaultmidpunct}
{\mcitedefaultendpunct}{\mcitedefaultseppunct}\relax
\EndOfBibitem
\bibitem[Nooijen and Bartlett(1995)Nooijen, and
  Bartlett]{Marcel_EA_EOM_CC_jcp_1995}
Nooijen,~M.; Bartlett,~R.~J. \emph{J. Chem. Phys.} \textbf{1995}, \emph{102},
  3629--3647\relax
\mciteBstWouldAddEndPuncttrue
\mciteSetBstMidEndSepPunct{\mcitedefaultmidpunct}
{\mcitedefaultendpunct}{\mcitedefaultseppunct}\relax
\EndOfBibitem
\bibitem[Nooijen and Bartlett(1997)Nooijen, and
  Bartlett]{Marcel_SimTrans_EOM_CC_IP_EA_EE_jcp_1997}
Nooijen,~M.; Bartlett,~R.~J. \emph{J. Chem. Phys.} \textbf{1997}, \emph{106},
  6449--6455\relax
\mciteBstWouldAddEndPuncttrue
\mciteSetBstMidEndSepPunct{\mcitedefaultmidpunct}
{\mcitedefaultendpunct}{\mcitedefaultseppunct}\relax
\EndOfBibitem
\bibitem[Musial et~al.(2003)Musial, Kucharski, and Bartlett]{Musial_IP}
Musial,~M.; Kucharski,~S.~A.; Bartlett,~R.~J. \emph{The Journal of Chemical
  Physics} \textbf{2003}, \emph{118}, 1128--1136\relax
\mciteBstWouldAddEndPuncttrue
\mciteSetBstMidEndSepPunct{\mcitedefaultmidpunct}
{\mcitedefaultendpunct}{\mcitedefaultseppunct}\relax
\EndOfBibitem
\bibitem[Musial and Bartlett(2003)Musial, and Bartlett]{Musial_EA}
Musial,~M.; Bartlett,~R.~J. \emph{J. Chem. Phys.} \textbf{2003}, \emph{119},
  1901--1908\relax
\mciteBstWouldAddEndPuncttrue
\mciteSetBstMidEndSepPunct{\mcitedefaultmidpunct}
{\mcitedefaultendpunct}{\mcitedefaultseppunct}\relax
\EndOfBibitem
\bibitem[Musial and Bartlett(2004)Musial, and Bartlett]{Musial_organic}
Musial,~M.; Bartlett,~R.~J. \emph{Chem. Phys. Lett.} \textbf{2004}, \emph{384},
  210 -- 214\relax
\mciteBstWouldAddEndPuncttrue
\mciteSetBstMidEndSepPunct{\mcitedefaultmidpunct}
{\mcitedefaultendpunct}{\mcitedefaultseppunct}\relax
\EndOfBibitem
\bibitem[Kamiya and Hirata(2007)Kamiya, and
  Hirata]{Kamiya_Hirata_EA_EOM_CC_jcp_2007}
Kamiya,~M.; Hirata,~S. \emph{J. Chem. Phys.} \textbf{2007}, \emph{126},
  --\relax
\mciteBstWouldAddEndPuncttrue
\mciteSetBstMidEndSepPunct{\mcitedefaultmidpunct}
{\mcitedefaultendpunct}{\mcitedefaultseppunct}\relax
\EndOfBibitem
\bibitem[Musial and Bartlett(2007)Musial, and Bartlett]{Musial}
Musial,~M.; Bartlett,~R.~J. \emph{J. Chem. Phys.} \textbf{2007}, \emph{127},
  --\relax
\mciteBstWouldAddEndPuncttrue
\mciteSetBstMidEndSepPunct{\mcitedefaultmidpunct}
{\mcitedefaultendpunct}{\mcitedefaultseppunct}\relax
\EndOfBibitem
\bibitem[Nooijen and Snijders(1992)Nooijen, and
  Snijders]{Marcel_CC_GreensFunction_ijqc_1992}
Nooijen,~M.; Snijders,~J.~G. \emph{Int. J. Quantum Chem.} \textbf{1992},
  \emph{44}, 55--83\relax
\mciteBstWouldAddEndPuncttrue
\mciteSetBstMidEndSepPunct{\mcitedefaultmidpunct}
{\mcitedefaultendpunct}{\mcitedefaultseppunct}\relax
\EndOfBibitem
\bibitem[Nooijen and Snijders(1993)Nooijen, and
  Snijders]{Marcel_CC_GreensFunction_ijqc_1993}
Nooijen,~M.; Snijders,~J.~G. \emph{Int. J. Quantum Chem.} \textbf{1993},
  \emph{48}, 15--48\relax
\mciteBstWouldAddEndPuncttrue
\mciteSetBstMidEndSepPunct{\mcitedefaultmidpunct}
{\mcitedefaultendpunct}{\mcitedefaultseppunct}\relax
\EndOfBibitem
\bibitem[Nooijen and Snijders(1995)Nooijen, and
  Snijders]{Marcel_gf2_CC_IP_jcp_1995}
Nooijen,~M.; Snijders,~J.~G. \emph{The Journal of Chemical Physics}
  \textbf{1995}, \emph{102}, 1681--1688\relax
\mciteBstWouldAddEndPuncttrue
\mciteSetBstMidEndSepPunct{\mcitedefaultmidpunct}
{\mcitedefaultendpunct}{\mcitedefaultseppunct}\relax
\EndOfBibitem
\bibitem[Schirmer(1982)]{ADCn}
Schirmer,~J. \emph{Phys. Rev. A} \textbf{1982}, \emph{26}, 2395--2416\relax
\mciteBstWouldAddEndPuncttrue
\mciteSetBstMidEndSepPunct{\mcitedefaultmidpunct}
{\mcitedefaultendpunct}{\mcitedefaultseppunct}\relax
\EndOfBibitem
\bibitem[Tarantelli and Cederbaum(1989)Tarantelli, and Cederbaum]{ADC_pra_1989}
Tarantelli,~A.; Cederbaum,~L.~S. \emph{Phys. Rev. A} \textbf{1989}, \emph{39},
  1656--1664\relax
\mciteBstWouldAddEndPuncttrue
\mciteSetBstMidEndSepPunct{\mcitedefaultmidpunct}
{\mcitedefaultendpunct}{\mcitedefaultseppunct}\relax
\EndOfBibitem
\bibitem[Flores-Moreno et~al.(2010)Flores-Moreno, Melin, Dolgounitcheva,
  Zakrzewski, and Ortiz]{OrtizThreeApproximations_ijqc_2010}
Flores-Moreno,~R.; Melin,~J.; Dolgounitcheva,~O.; Zakrzewski,~V.~G.;
  Ortiz,~J.~V. \emph{Int. J. Quantum Chem.} \textbf{2010}, \emph{110},
  706--715\relax
\mciteBstWouldAddEndPuncttrue
\mciteSetBstMidEndSepPunct{\mcitedefaultmidpunct}
{\mcitedefaultendpunct}{\mcitedefaultseppunct}\relax
\EndOfBibitem
\bibitem[Cederbaum(1973)]{DirectIPs_Cederbaum_1973}
Cederbaum,~L. \emph{Theoretica chimica acta} \textbf{1973}, \emph{31},
  239--260\relax
\mciteBstWouldAddEndPuncttrue
\mciteSetBstMidEndSepPunct{\mcitedefaultmidpunct}
{\mcitedefaultendpunct}{\mcitedefaultseppunct}\relax
\EndOfBibitem
\bibitem[Ortiz(1988)]{Ortiz_Partial4thOrder_jcp_1988}
Ortiz,~J.~V. \emph{J. Chem. Phys.} \textbf{1988}, \emph{89}, 6348--6352\relax
\mciteBstWouldAddEndPuncttrue
\mciteSetBstMidEndSepPunct{\mcitedefaultmidpunct}
{\mcitedefaultendpunct}{\mcitedefaultseppunct}\relax
\EndOfBibitem
\bibitem[Ortiz(1989)]{Ortiz_NonDiagPartial4th_ijqc_1989}
Ortiz,~J.~V. \emph{Int. J. Quantum Chem.} \textbf{1989}, \emph{36},
  321--332\relax
\mciteBstWouldAddEndPuncttrue
\mciteSetBstMidEndSepPunct{\mcitedefaultmidpunct}
{\mcitedefaultendpunct}{\mcitedefaultseppunct}\relax
\EndOfBibitem
\bibitem[Ortiz(1996)]{Ortiz_Partial3rdOrder_jcp_1996}
Ortiz,~J.~V. \emph{The Journal of Chemical Physics} \textbf{1996}, \emph{104},
  7599--7605\relax
\mciteBstWouldAddEndPuncttrue
\mciteSetBstMidEndSepPunct{\mcitedefaultmidpunct}
{\mcitedefaultendpunct}{\mcitedefaultseppunct}\relax
\EndOfBibitem
\bibitem[Ortiz(1998)]{Ortiz_NonDiagRenorm_3rdOrder_jcp_1998}
Ortiz,~J.~V. \emph{The Journal of Chemical Physics} \textbf{1998}, \emph{108},
  1008--1014\relax
\mciteBstWouldAddEndPuncttrue
\mciteSetBstMidEndSepPunct{\mcitedefaultmidpunct}
{\mcitedefaultendpunct}{\mcitedefaultseppunct}\relax
\EndOfBibitem
\bibitem[Deleuze et~al.(1999)Deleuze, Giuffreda, Fran√ßois, and
  Cederbaum]{CarbonCluster_ADC_jcp_1999}
Deleuze,~M.~S.; Giuffreda,~M.~G.; Fran√ßois,~J.-P.; Cederbaum,~L.~S.
  \emph{J. Chem. Phys.} \textbf{1999}, \emph{111}, 5851--5865\relax
\mciteBstWouldAddEndPuncttrue
\mciteSetBstMidEndSepPunct{\mcitedefaultmidpunct}
{\mcitedefaultendpunct}{\mcitedefaultseppunct}\relax
\EndOfBibitem
\bibitem[Seabra et~al.(2004)Seabra, Kaplan, Zakrzewski, and
  Ortiz]{Ortiz_EPT_jcp_2004}
Seabra,~G.~M.; Kaplan,~I.~G.; Zakrzewski,~V.~G.; Ortiz,~J.~V. \emph{The Journal
  of Chemical Physics} \textbf{2004}, \emph{121}, 4143--4155\relax
\mciteBstWouldAddEndPuncttrue
\mciteSetBstMidEndSepPunct{\mcitedefaultmidpunct}
{\mcitedefaultendpunct}{\mcitedefaultseppunct}\relax
\EndOfBibitem
\bibitem[Trofimov and Schirmer(2005)Trofimov, and
  Schirmer]{NonDysonADC_jcp_2005}
Trofimov,~A.~B.; Schirmer,~J. \emph{The Journal of Chemical Physics}
  \textbf{2005}, \emph{123}, --\relax
\mciteBstWouldAddEndPuncttrue
\mciteSetBstMidEndSepPunct{\mcitedefaultmidpunct}
{\mcitedefaultendpunct}{\mcitedefaultseppunct}\relax
\EndOfBibitem
\bibitem[Starcke et~al.(2006)Starcke, Wormit, Schirmer, and
  Dreuw]{HowMuchDoublePolyene_ADC_2006}
Starcke,~J.~H.; Wormit,~M.; Schirmer,~J.; Dreuw,~A. \emph{Chem. Phys.}
  \textbf{2006}, \emph{329}, 39 -- 49, Electron Correlation and Multimode
  Dynamics in Molecules (in honour of Lorenz S. Cederbaum)\relax
\mciteBstWouldAddEndPuncttrue
\mciteSetBstMidEndSepPunct{\mcitedefaultmidpunct}
{\mcitedefaultendpunct}{\mcitedefaultseppunct}\relax
\EndOfBibitem
\bibitem[M\"uller and Cederbaum(2006)M\"uller, and
  Cederbaum]{Cederbaum_Water_DoubleIP_jcp_2006}
M\"uller,~I.~B.; Cederbaum,~L.~S. \emph{J. Chem. Phys.} \textbf{2006},
  \emph{125}, --\relax
\mciteBstWouldAddEndPuncttrue
\mciteSetBstMidEndSepPunct{\mcitedefaultmidpunct}
{\mcitedefaultendpunct}{\mcitedefaultseppunct}\relax
\EndOfBibitem
\bibitem[Godby et~al.(1986)Godby, Schl\"uter, and
  Sham]{AccXCPotSi_Discontinuity_Godby_prl_1986}
Godby,~R.~W.; Schl\"uter,~M.; Sham,~L.~J. \emph{Phys. Rev. Lett.}
  \textbf{1986}, \emph{56}, 2415--2418\relax
\mciteBstWouldAddEndPuncttrue
\mciteSetBstMidEndSepPunct{\mcitedefaultmidpunct}
{\mcitedefaultendpunct}{\mcitedefaultseppunct}\relax
\EndOfBibitem
\bibitem[Gr\"uning et~al.(2006)Gr\"uning, Marini, and
  Rubio]{DFTMBPT_BandGap_Gruning_jcp_2006}
Gr\"uning,~M.; Marini,~A.; Rubio,~A. \emph{J. Chem. Phys.} \textbf{2006},
  \emph{124}, --\relax
\mciteBstWouldAddEndPuncttrue
\mciteSetBstMidEndSepPunct{\mcitedefaultmidpunct}
{\mcitedefaultendpunct}{\mcitedefaultseppunct}\relax
\EndOfBibitem
\bibitem[Perdew and Zunger(1981)Perdew, and Zunger]{SIEcPerdewZunger1981}
Perdew,~J.~P.; Zunger,~A. \emph{Phys. Rev. B} \textbf{1981}, \emph{23},
  5048--5079\relax
\mciteBstWouldAddEndPuncttrue
\mciteSetBstMidEndSepPunct{\mcitedefaultmidpunct}
{\mcitedefaultendpunct}{\mcitedefaultseppunct}\relax
\EndOfBibitem
\bibitem[Perdew et~al.(1982)Perdew, Parr, Levy, and Balduz~Jr.]{Perdew1982}
Perdew,~J.~P.; Parr,~R.~G.; Levy,~M.; Balduz~Jr.,~J.~L. \emph{Phys. Rev. Lett.}
  \textbf{1982}, \emph{49}, 1691\relax
\mciteBstWouldAddEndPuncttrue
\mciteSetBstMidEndSepPunct{\mcitedefaultmidpunct}
{\mcitedefaultendpunct}{\mcitedefaultseppunct}\relax
\EndOfBibitem
\bibitem[Ruzsinszky et~al.(2006)Ruzsinszky, Perdew, Csonka, Vydrov, and
  Scuseria]{SpuriousFrac2006}
Ruzsinszky,~A.; Perdew,~J.~P.; Csonka,~G.~I.; Vydrov,~O.~A.; Scuseria,~G.~E.
  \emph{J. Chem. Phys.} \textbf{2006}, \emph{125}, 194112\relax
\mciteBstWouldAddEndPuncttrue
\mciteSetBstMidEndSepPunct{\mcitedefaultmidpunct}
{\mcitedefaultendpunct}{\mcitedefaultseppunct}\relax
\EndOfBibitem
\bibitem[S\`{a}nchez et~al.(2006)S\`{a}nchez, Cohen, and
  Yang]{ManyeSIEYang2006}
S\`{a}nchez,~P.~M.; Cohen,~A.~J.; Yang,~W. \emph{J. Chem. Phys.} \textbf{2006},
  \emph{125}, 201102\relax
\mciteBstWouldAddEndPuncttrue
\mciteSetBstMidEndSepPunct{\mcitedefaultmidpunct}
{\mcitedefaultendpunct}{\mcitedefaultseppunct}\relax
\EndOfBibitem
\bibitem[Cohen et~al.(2008)Cohen, Mori-S\'anchez, and
  Yang]{FractionalBandGap_Cohen_prb_2008}
Cohen,~A.~J.; Mori-S\'anchez,~P.; Yang,~W. \emph{Phys. Rev. B} \textbf{2008},
  \emph{77}, 115123\relax
\mciteBstWouldAddEndPuncttrue
\mciteSetBstMidEndSepPunct{\mcitedefaultmidpunct}
{\mcitedefaultendpunct}{\mcitedefaultseppunct}\relax
\EndOfBibitem
\bibitem[Moussa et~al.(2012)Moussa, Schultz, and
  Chelikowsky]{HSE_ParameterSpace_Moussa_jcp_2012}
Moussa,~J.~E.; Schultz,~P.~A.; Chelikowsky,~J.~R. \emph{J. Chem. Phys.}
  \textbf{2012}, \emph{136}, --\relax
\mciteBstWouldAddEndPuncttrue
\mciteSetBstMidEndSepPunct{\mcitedefaultmidpunct}
{\mcitedefaultendpunct}{\mcitedefaultseppunct}\relax
\EndOfBibitem
\bibitem[Seidl et~al.(1996)Seidl, G\"orling, Vogl, Majewski, and
  Levy]{GeneralizedKohnSham_BandGap_Levy_prb_1996}
Seidl,~A.; G\"orling,~A.; Vogl,~P.; Majewski,~J.~A.; Levy,~M. \emph{Phys. Rev.
  B} \textbf{1996}, \emph{53}, 3764--3774\relax
\mciteBstWouldAddEndPuncttrue
\mciteSetBstMidEndSepPunct{\mcitedefaultmidpunct}
{\mcitedefaultendpunct}{\mcitedefaultseppunct}\relax
\EndOfBibitem
\bibitem[Heyd et~al.(2003)Heyd, Scuseria, and Ernzerhof]{Heyd:2003db}
Heyd,~J.; Scuseria,~G.; Ernzerhof,~M. \emph{J. Chem. Phys.} \textbf{2003},
  \emph{118}, 8207--8215\relax
\mciteBstWouldAddEndPuncttrue
\mciteSetBstMidEndSepPunct{\mcitedefaultmidpunct}
{\mcitedefaultendpunct}{\mcitedefaultseppunct}\relax
\EndOfBibitem
\bibitem[Heyd and Scuseria(2004)Heyd, and Scuseria]{Heyd:2004ud}
Heyd,~J.; Scuseria,~G. \emph{J. Chem. Phys.} \textbf{2004}, \emph{120},
  7274--7280\relax
\mciteBstWouldAddEndPuncttrue
\mciteSetBstMidEndSepPunct{\mcitedefaultmidpunct}
{\mcitedefaultendpunct}{\mcitedefaultseppunct}\relax
\EndOfBibitem
\bibitem[Heyd et~al.(2005)Heyd, Peralta, Scuseria, and
  Martin]{EnergyBandGaps_HSE_Heyd_Peralta_jcp_2005}
Heyd,~J.; Peralta,~J.~E.; Scuseria,~G.~E.; Martin,~R.~L. \emph{J. Chem. Phys.}
  \textbf{2005}, \emph{123}, --\relax
\mciteBstWouldAddEndPuncttrue
\mciteSetBstMidEndSepPunct{\mcitedefaultmidpunct}
{\mcitedefaultendpunct}{\mcitedefaultseppunct}\relax
\EndOfBibitem
\bibitem[Brothers et~al.(2008)Brothers, Izmaylov, Normand, Barone, and
  Scuseria]{BandGapsHSE_vero_barone_jcp_2008}
Brothers,~E.~N.; Izmaylov,~A.~F.; Normand,~J.~O.; Barone,~V.; Scuseria,~G.~E.
  \emph{J. Chem. Phys.} \textbf{2008}, \emph{129}, --\relax
\mciteBstWouldAddEndPuncttrue
\mciteSetBstMidEndSepPunct{\mcitedefaultmidpunct}
{\mcitedefaultendpunct}{\mcitedefaultseppunct}\relax
\EndOfBibitem
\bibitem[Henderson et~al.(2011)Henderson, Paier, and
  Scuseria]{AccurateSolidsWithHSE_Henderson_pss_2011}
Henderson,~T.~M.; Paier,~J.; Scuseria,~G.~E. \emph{physica status solidi b}
  \textbf{2011}, \emph{248}, 767--774\relax
\mciteBstWouldAddEndPuncttrue
\mciteSetBstMidEndSepPunct{\mcitedefaultmidpunct}
{\mcitedefaultendpunct}{\mcitedefaultseppunct}\relax
\EndOfBibitem
\bibitem[Hybertsen and Louie(1986)Hybertsen, and Louie]{Hybertsen_GW_prb_1986}
Hybertsen,~M.~S.; Louie,~S.~G. \emph{Phys. Rev. B} \textbf{1986}, \emph{34},
  5390--5413\relax
\mciteBstWouldAddEndPuncttrue
\mciteSetBstMidEndSepPunct{\mcitedefaultmidpunct}
{\mcitedefaultendpunct}{\mcitedefaultseppunct}\relax
\EndOfBibitem
\bibitem[Onida et~al.(2002)Onida, Reining, and Rubio]{GW_Rubio_RevModPhys_2002}
Onida,~G.; Reining,~L.; Rubio,~A. \emph{Rev. Mod. Phys.} \textbf{2002},
  \emph{74}, 601--659\relax
\mciteBstWouldAddEndPuncttrue
\mciteSetBstMidEndSepPunct{\mcitedefaultmidpunct}
{\mcitedefaultendpunct}{\mcitedefaultseppunct}\relax
\EndOfBibitem
\bibitem[Fuchs et~al.(2007)Fuchs, Furthm\"uller, Bechstedt, Shishkin, and
  Kresse]{GWStartingPoint_Kresse_prv_2007}
Fuchs,~F.; Furthm\"uller,~J.; Bechstedt,~F.; Shishkin,~M.; Kresse,~G.
  \emph{Phys. Rev. B} \textbf{2007}, \emph{76}, 115109\relax
\mciteBstWouldAddEndPuncttrue
\mciteSetBstMidEndSepPunct{\mcitedefaultmidpunct}
{\mcitedefaultendpunct}{\mcitedefaultseppunct}\relax
\EndOfBibitem
\bibitem[Blase et~al.(2011)Blase, Attaccalite, and
  Olevano]{GWStartingPoint_prb_2011}
Blase,~X.; Attaccalite,~C.; Olevano,~V. \emph{Phys. Rev. B} \textbf{2011},
  \emph{83}, 115103\relax
\mciteBstWouldAddEndPuncttrue
\mciteSetBstMidEndSepPunct{\mcitedefaultmidpunct}
{\mcitedefaultendpunct}{\mcitedefaultseppunct}\relax
\EndOfBibitem
\bibitem[Marom et~al.(2011)Marom, Moussa, Ren, Tkatchenko, and
  Chelikowsky]{GWStartingPoint_TiO2_Gap_Moussa_prb_2011}
Marom,~N.; Moussa,~J.~E.; Ren,~X.; Tkatchenko,~A.; Chelikowsky,~J.~R.
  \emph{Phys. Rev. B} \textbf{2011}, \emph{84}, 245115\relax
\mciteBstWouldAddEndPuncttrue
\mciteSetBstMidEndSepPunct{\mcitedefaultmidpunct}
{\mcitedefaultendpunct}{\mcitedefaultseppunct}\relax
\EndOfBibitem
\bibitem[Marom et~al.(2011)Marom, Ren, Moussa, Chelikowsky, and
  Kronik]{GWStartingPoint_Kronik_prb_2011}
Marom,~N.; Ren,~X.; Moussa,~J.~E.; Chelikowsky,~J.~R.; Kronik,~L. \emph{Phys.
  Rev. B} \textbf{2011}, \emph{84}, 195143\relax
\mciteBstWouldAddEndPuncttrue
\mciteSetBstMidEndSepPunct{\mcitedefaultmidpunct}
{\mcitedefaultendpunct}{\mcitedefaultseppunct}\relax
\EndOfBibitem
\bibitem[Liao and Carter(2011)Liao, and
  Carter]{oneshotGWReference_Carter_pccp_2011}
Liao,~P.; Carter,~E.~A. \emph{Phys. Chem. Chem. Phys.} \textbf{2011},
  \emph{13}, 15189--15199\relax
\mciteBstWouldAddEndPuncttrue
\mciteSetBstMidEndSepPunct{\mcitedefaultmidpunct}
{\mcitedefaultendpunct}{\mcitedefaultseppunct}\relax
\EndOfBibitem
\bibitem[Toroker et~al.(2011)Toroker, Kanan, Alidoust, Isseroff, Liao, and
  Carter]{DFT_MetalOxide_BlasphemyPaper_pccp_2011}
Toroker,~M.~C.; Kanan,~D.~K.; Alidoust,~N.; Isseroff,~L.~Y.; Liao,~P.;
  Carter,~E.~A. \emph{Phys. Chem. Chem. Phys.} \textbf{2011}, \emph{13},
  16644--16654\relax
\mciteBstWouldAddEndPuncttrue
\mciteSetBstMidEndSepPunct{\mcitedefaultmidpunct}
{\mcitedefaultendpunct}{\mcitedefaultseppunct}\relax
\EndOfBibitem
\bibitem[Isseroff and Carter(2012)Isseroff, and
  Carter]{oneshotGWReference_Carter_prb_2012}
Isseroff,~L.~Y.; Carter,~E.~A. \emph{Phys. Rev. B} \textbf{2012}, \emph{85},
  235142\relax
\mciteBstWouldAddEndPuncttrue
\mciteSetBstMidEndSepPunct{\mcitedefaultmidpunct}
{\mcitedefaultendpunct}{\mcitedefaultseppunct}\relax
\EndOfBibitem
\bibitem[Marom et~al.(2012)Marom, Caruso, Ren, Hofmann, K\"orzd\"orfer,
  Chelikowsky, Rubio, Scheffler, and Rinke]{GWBenchmark_prb_2012}
Marom,~N.; Caruso,~F.; Ren,~X.; Hofmann,~O.~T.; K\"orzd\"orfer,~T.;
  Chelikowsky,~J.~R.; Rubio,~A.; Scheffler,~M.; Rinke,~P. \emph{Phys. Rev. B}
  \textbf{2012}, \emph{86}, 245127\relax
\mciteBstWouldAddEndPuncttrue
\mciteSetBstMidEndSepPunct{\mcitedefaultmidpunct}
{\mcitedefaultendpunct}{\mcitedefaultseppunct}\relax
\EndOfBibitem
\bibitem[K\"orzd\"orfer and Marom(2012)K\"orzd\"orfer, and
  Marom]{GWStartingPoint_prb_2012}
K\"orzd\"orfer,~T.; Marom,~N. \emph{Phys. Rev. B} \textbf{2012}, \emph{86},
  041110\relax
\mciteBstWouldAddEndPuncttrue
\mciteSetBstMidEndSepPunct{\mcitedefaultmidpunct}
{\mcitedefaultendpunct}{\mcitedefaultseppunct}\relax
\EndOfBibitem
\bibitem[Bruneval and Marques(2013)Bruneval, and
  Marques]{GWStartingPoint_jctc_2013}
Bruneval,~F.; Marques,~M. A.~L. \emph{J. Chem. Theory Comput.} \textbf{2013},
  \emph{9}, 324--329\relax
\mciteBstWouldAddEndPuncttrue
\mciteSetBstMidEndSepPunct{\mcitedefaultmidpunct}
{\mcitedefaultendpunct}{\mcitedefaultseppunct}\relax
\EndOfBibitem
\bibitem[Perdew et~al.(1996)Perdew, Burke, and Ernzerhof]{PBE}
Perdew,~J.~P.; Burke,~K.; Ernzerhof,~M. \emph{Phys. Rev. Lett.} \textbf{1996},
  \emph{77}, 3865--3868\relax
\mciteBstWouldAddEndPuncttrue
\mciteSetBstMidEndSepPunct{\mcitedefaultmidpunct}
{\mcitedefaultendpunct}{\mcitedefaultseppunct}\relax
\EndOfBibitem
\bibitem[Ernzerhof and Scuseria(1999)Ernzerhof, and Scuseria]{PBE1}
Ernzerhof,~M.; Scuseria,~G.~E. \emph{J. Chem. Phys.} \textbf{1999}, \emph{110},
  5029--5036\relax
\mciteBstWouldAddEndPuncttrue
\mciteSetBstMidEndSepPunct{\mcitedefaultmidpunct}
{\mcitedefaultendpunct}{\mcitedefaultseppunct}\relax
\EndOfBibitem
\bibitem[Adamo and Barone(1999)Adamo, and Barone]{PBE0}
Adamo,~C.; Barone,~V. \emph{J. Chem. Phys.} \textbf{1999}, \emph{110},
  6158--6170\relax
\mciteBstWouldAddEndPuncttrue
\mciteSetBstMidEndSepPunct{\mcitedefaultmidpunct}
{\mcitedefaultendpunct}{\mcitedefaultseppunct}\relax
\EndOfBibitem
\bibitem[Holm and von Barth(1998)Holm, and von
  Barth]{scGW_Holm_vonBarth_PRB_1998}
Holm,~B.; von Barth,~U. \emph{Phys. Rev. B} \textbf{1998}, \emph{57},
  2108--2117\relax
\mciteBstWouldAddEndPuncttrue
\mciteSetBstMidEndSepPunct{\mcitedefaultmidpunct}
{\mcitedefaultendpunct}{\mcitedefaultseppunct}\relax
\EndOfBibitem
\bibitem[Sch\"one and Eguiluz(1998)Sch\"one, and
  Eguiluz]{scGW_Semiconductors_prl_1998}
Sch\"one,~W.-D.; Eguiluz,~A.~G. \emph{Phys. Rev. Lett.} \textbf{1998},
  \emph{81}, 1662--1665\relax
\mciteBstWouldAddEndPuncttrue
\mciteSetBstMidEndSepPunct{\mcitedefaultmidpunct}
{\mcitedefaultendpunct}{\mcitedefaultseppunct}\relax
\EndOfBibitem
\bibitem[Faber et~al.(2011)Faber, Attaccalite, Olevano, Runge, and
  Blase]{GW_DNA_selfconsistent_prb_2011}
Faber,~C.; Attaccalite,~C.; Olevano,~V.; Runge,~E.; Blase,~X. \emph{Phys. Rev.
  B} \textbf{2011}, \emph{83}, 115123\relax
\mciteBstWouldAddEndPuncttrue
\mciteSetBstMidEndSepPunct{\mcitedefaultmidpunct}
{\mcitedefaultendpunct}{\mcitedefaultseppunct}\relax
\EndOfBibitem
\bibitem[Strange et~al.(2011)Strange, Rostgaard, H\"akkinen, and
  Thygesen]{scGW_transport_prb_2011}
Strange,~M.; Rostgaard,~C.; H\"akkinen,~H.; Thygesen,~K.~S. \emph{Phys. Rev. B}
  \textbf{2011}, \emph{83}, 115108\relax
\mciteBstWouldAddEndPuncttrue
\mciteSetBstMidEndSepPunct{\mcitedefaultmidpunct}
{\mcitedefaultendpunct}{\mcitedefaultseppunct}\relax
\EndOfBibitem
\bibitem[Caruso et~al.(2012)Caruso, Rinke, Ren, Scheffler, and
  Rubio]{scGW_Unified_Scheffler_prb_2012}
Caruso,~F.; Rinke,~P.; Ren,~X.; Scheffler,~M.; Rubio,~A. \emph{Phys. Rev. B}
  \textbf{2012}, \emph{86}, 081102\relax
\mciteBstWouldAddEndPuncttrue
\mciteSetBstMidEndSepPunct{\mcitedefaultmidpunct}
{\mcitedefaultendpunct}{\mcitedefaultseppunct}\relax
\EndOfBibitem
\bibitem[Mattuck(1976)]{Mattuck_Feynman}
Mattuck,~R. \emph{A Guide to Feynman Diagrams in the Many-body Problem}; Dover
  Books on Physics Series; Dover Publications, Incorporated, 1976\relax
\mciteBstWouldAddEndPuncttrue
\mciteSetBstMidEndSepPunct{\mcitedefaultmidpunct}
{\mcitedefaultendpunct}{\mcitedefaultseppunct}\relax
\EndOfBibitem
\bibitem[Fetter and Walecka(2003)Fetter, and Walecka]{fetter2003quantum}
Fetter,~A.; Walecka,~J. \emph{Quantum Theory of Many-particle Systems}; Dover
  Books on Physics; Dover Publications, 2003; p~65\relax
\mciteBstWouldAddEndPuncttrue
\mciteSetBstMidEndSepPunct{\mcitedefaultmidpunct}
{\mcitedefaultendpunct}{\mcitedefaultseppunct}\relax
\EndOfBibitem
\bibitem[Albuquerque et~al.(2007)Albuquerque, Alet, Corboz, Dayal, Feiguin,
  Fuchs, Gamper, Gull, G{\"u}rtler, Honecker, Igarashi, K\"{o}rner,
  Kozhevnikov, L\"{a}uchli, Manmana, Matsumoto, McCulloch, Michel, Noack,
  Paw{\l}owski, Pollet, Pruschke, Schollw\"{o}ck, Todo, Trebst, Troyer, Werner,
  and Wessel]{ALPS}
Albuquerque,~A. et~al.  \emph{J. Magn. Magn. Mater.} \textbf{2007}, \emph{310},
  1187 -- 1193, Proceedings of the 17th International Conference on Magnetism
  The International Conference on Magnetism\relax
\mciteBstWouldAddEndPuncttrue
\mciteSetBstMidEndSepPunct{\mcitedefaultmidpunct}
{\mcitedefaultendpunct}{\mcitedefaultseppunct}\relax
\EndOfBibitem
\bibitem[Aidas et~al.(2013)Aidas, Angeli, Bak, Bakken, Bast, Boman,
  Christiansen, Cimiraglia, Coriani, Dahle, Dalskov, Ekstr\"{o}m, Enevoldsen,
  Eriksen, Ettenhuber, Fern{\'a}ndez, Ferrighi, Fliegl, Frediani, Hald,
  Halkier, H\"{a}ttig, Heiberg, Helgaker, Hennum, Hettema, Hjerten{\ae}s,
  H{\o}st, H{\o}yvik, Iozzi, Jans\'{i}k, Jensen, Jonsson, J{\o}rgensen,
  Kauczor, Kirpekar, Kj{\ae}rgaard, Klopper, Knecht, Kobayashi, Koch, Kongsted,
  Krapp, Kristensen, Ligabue, Lutn{\ae}s, Melo, Mikkelsen, Myhre, Neiss,
  Nielsen, Norman, Olsen, Olsen, Osted, Packer, Pawlowski, Pedersen, Provasi,
  Reine, Rinkevicius, Ruden, Ruud, Rybkin, Sa{\l}ek, Samson, de~Mer\'{a}s,
  Saue, Sauer, Schimmelpfennig, Sneskov, Steindal, Sylvester-Hvid, Taylor,
  Teale, Tellgren, Tew, Thorvaldsen, Th{\o}gersen, Vahtras, Watson, Wilson,
  Ziolkowski, and √Ögren]{Dalton}
Aidas,~K. et~al.  \emph{Wiley Interdisciplinary Reviews: Computational
  Molecular Science} \textbf{2013}, n/a--n/a\relax
\mciteBstWouldAddEndPuncttrue
\mciteSetBstMidEndSepPunct{\mcitedefaultmidpunct}
{\mcitedefaultendpunct}{\mcitedefaultseppunct}\relax
\EndOfBibitem
\bibitem[Johnson(2011)]{geometry}
Johnson,~R.~D. {NIST Computational Chemistry Comparison and Benchmark
  Database}. 2011\relax
\mciteBstWouldAddEndPuncttrue
\mciteSetBstMidEndSepPunct{\mcitedefaultmidpunct}
{\mcitedefaultendpunct}{\mcitedefaultseppunct}\relax
\EndOfBibitem
\bibitem[Dunning(1989)]{basis1}
Dunning,~T.~H. \emph{The Journal of Chemical Physics} \textbf{1989}, \emph{90},
  1007--1023\relax
\mciteBstWouldAddEndPuncttrue
\mciteSetBstMidEndSepPunct{\mcitedefaultmidpunct}
{\mcitedefaultendpunct}{\mcitedefaultseppunct}\relax
\EndOfBibitem
\bibitem[Woon and Dunning(1993)Woon, and Dunning]{basis2}
Woon,~D.~E.; Dunning,~T.~H. \emph{The Journal of Chemical Physics}
  \textbf{1993}, \emph{98}, 1358--1371\relax
\mciteBstWouldAddEndPuncttrue
\mciteSetBstMidEndSepPunct{\mcitedefaultmidpunct}
{\mcitedefaultendpunct}{\mcitedefaultseppunct}\relax
\EndOfBibitem
\bibitem[Frisch et~al.(2009)Frisch, Trucks, Schlegel, Scuseria, Robb,
  Cheeseman, Scalmani, Barone, Mennucci, Petersson, Nakatsuji, Caricato, Li,
  Hratchian, Izmaylov, Bloino, Zheng, Sonnenberg, Hada, Ehara, Toyota, Fukuda,
  Hasegawa, Ishida, Nakajima, Honda, Kitao, Nakai, Vreven, Montgomery, Peralta,
  Ogliaro, Bearpark, Heyd, Brothers, Kudin, Staroverov, Kobayashi, Normand,
  Raghavachari, Rendell, Burant, Iyengar, Tomasi, Cossi, Rega, Millam, Klene,
  Knox, Cross, Bakken, Adamo, Jaramillo, Gomperts, Stratmann, Yazyev, Austin,
  Cammi, Pomelli, Ochterski, Martin, Morokuma, Zakrzewski, Voth, Salvador,
  Dannenberg, Dapprich, Daniels, Farkas, Foresman, Ortiz, Cioslowski, and
  Fox]{g09}
Frisch,~M.~J. et~al.  Gaussian~09 {R}evision {A}.1. 2009; {G}aussian Inc.
  Wallingford CT\relax
\mciteBstWouldAddEndPuncttrue
\mciteSetBstMidEndSepPunct{\mcitedefaultmidpunct}
{\mcitedefaultendpunct}{\mcitedefaultseppunct}\relax
\EndOfBibitem
\end{mcitethebibliography}
\end{document}